\documentclass[letterpaper,doc,12pt,floatsintext,longtable]{apa6}

\newcommand{\papertitle}{Distilling Black-Box Machine Learning into a Small, Self-Explaining Language Model for Learning Analytics}   
\newcommand{\shortpapertitle}{Distilling Machine Learning into an LLM}
\newcommand{\authornames}{Chenguang Pan*, Airui Meng,and Youmi Suk}
\newcommand{\institution}{{Department of Human Development, Teachers College Columbia University}}
\newcommand{\authoremails}{\{cp3280, am6817, ysuk\}@tc.columbia.edu}
\newcommand{\paperdate}{August 21, 2026}
\newcommand{\correspondingauthor}{Chenguang Pan}
\newcommand{\correspondingemail}{cp3280@tc.columbia.edu}
\newcommand{\abstracttext}{Learning analytics increasingly relies on flexible machine learning (ML), but the model opacity and the burden of deployment prevent these tools from reaching educational practice. We propose a two-stage fine-tuning pipeline that distills a fitted black-box estimator and its post hoc interpretation (the mentor) into a small, open-weight large language model (LLM; the mentee) that returns an individual-level estimate and explains in natural language. The design is estimator-agnostic and paired with a faithfulness-first evaluation framework that audits every narration against the attribution it claims to describe. We design a simulation study that separates distillation loss from estimator loss by comparing an oracle mentor with a realistic ML mentor. Given an oracle signal, distillation with a two-billion-parameter LLM model is nearly lossless in recovering the effect surface ($r > .90$), perfectly ranking the important variables, and citing no spurious covariate. Under a realistic estimator, almost all remaining error originates upstream. We find that fluency is no evidence of correctness since narration quality is independent of signal quality, and decision quality collapses toward the majority action in severely imbalanced settings. Applied to a nationally representative dataset, the pipeline recovers the finding that advanced mathematics coursework benefits students least likely to enroll in four-year college the most, with 98.8\% of narrations passing the audit and no fabricated quantities. The result is a single fine-tuned LLM that predicts and explains offline on a commodity laptop, so student records never leave the machine.}
\newcommand{\keywordlist}{knowledge distillation, large language models, explainable artificial intelligence, educational decision support, learning analytics}

\usepackage[american]{babel}
\usepackage[utf8]{inputenc}
\usepackage{amsmath}
\usepackage{amssymb}
\usepackage{bm}
\usepackage{graphicx}
\usepackage{xcolor}
\usepackage{colortbl}
\usepackage{caption}
\usepackage{subcaption}
\usepackage{setspace}
\usepackage{threeparttable}
\usepackage{endnotes}
\usepackage{xspace}
\usepackage{afterpage}
\usepackage{lscape}
\usepackage[colorlinks=true, allcolors=black]{hyperref}
\usepackage[normalem]{ulem}
\usepackage{csquotes}
\usepackage{dsfont}
\usepackage{booktabs}
\usepackage{multirow}
\usepackage{tabularx}
\usepackage{array}
\usepackage{arydshln}
\usepackage[noend]{algpseudocode}
\usepackage{algorithm}
\usepackage{listings}
\usepackage{bbm}

\usepackage{geometry}
\usepackage{tikz}
\usepackage{pgf}
\usetikzlibrary{
    matrix,
    positioning,
    calc,
    arrows,
    shapes.arrows,
    shapes.geometric,
    shapes.multipart,
    decorations.pathreplacing,
    decorations.pathmorphing,
    shapes,
    decorations,
    arrows.meta,
    fit
}

\definecolor{confcol}{HTML}{E6F1FB}\definecolor{confdraw}{HTML}{185FA5} 
\definecolor{predcol}{HTML}{E1F5EE}\definecolor{preddraw}{HTML}{0F6E56} 
\definecolor{neutcol}{HTML}{F1EFE8}\definecolor{neutdraw}{HTML}{5F5E5A} 
\definecolor{modcol}{HTML}{BA7517}                            

\tikzset{
  var/.style   = {circle, draw, minimum size=10mm, inner sep=1pt, font=\small,
                  line width=0.5pt},
  conf/.style  = {var, fill=confcol, draw=confdraw},
  pred/.style  = {var, fill=predcol, draw=preddraw},
  focal/.style = {var, fill=neutcol, draw=neutdraw, line width=0.9pt},
  noise/.style = {rectangle, rounded corners=3pt, draw=neutdraw, dashed,
                  fill=neutcol, font=\small, align=center, inner sep=5pt},
  mbadge/.style= {circle, fill=modcol, text=white, font=\bfseries\tiny,
                  inner sep=0.6pt},
  mod/.style   = {draw=modcol, line width=1pt, double, double distance=1.2pt,
                  label={[mbadge, label distance=-1.6mm]45:M}},
  cedge/.style = {-{Stealth[length=2mm]}, draw=confdraw, line width=0.5pt},
  pedge/.style = {-{Stealth[length=2mm]}, draw=preddraw, line width=0.5pt},
  tedge/.style = {-{Stealth[length=2mm]}, draw=neutdraw, line width=1pt},
  lkey/.style  = {circle, minimum size=4mm, inner sep=0pt, line width=0.5pt},
}

\usepackage{listings}
\graphicspath{{plot/}}

\usepackage[style=apa,sortcites=true,sorting=nyt,backend=biber]{biblatex}
\DeclareLanguageMapping{american}{american-apa}
\newcommand{\PreserveBackslash}[1]{\let\temp=\\#1\let\\=\temp}
\newcolumntype{C}[1]{>{\PreserveBackslash\centering}p{#1}}
\newcolumntype{L}[1]{>{\PreserveBackslash\raggedright}p{#1}}
\newcolumntype{R}[1]{>{\PreserveBackslash\raggedleft}p{#1}}

\title{\Large \papertitle}
\shorttitle{\shortpapertitle}
\author{\fontsize{13.5pt}{13pt}\selectfont \authornames \vspace{0.3em}}
\affiliation{\institution \\[0.3em]
\authoremails
\\[0.5em]
\paperdate}

\begin{document}

\maketitle

\begingroup
\renewcommand{\thefootnote}{\fnsymbol{footnote}}
\footnotetext[1]{Corresponding author: \correspondingauthor. Email: \correspondingemail.}
\endgroup

\vspace{-1em}
\begin{center}
\textbf{Abstract}
\end{center}

\noindent
\begin{minipage}{\textwidth}
\abstracttext

\vspace{0.75em}

\noindent\textit{Keywords:} \keywordlist
\end{minipage}

\vspace{1.5em}

\setcounter{secnumdepth}{3}


\section{Introduction}

Learning analytics and educational data mining increasingly rely on flexible Machine Learning (ML) models to study students' learning behaviors, enhance academic performance, and assist stakeholders in educational decision-making \parencite[]{hilbert2021machine,ifenthaler2017are,romero2020educational,pan2024examining}. Models such as random forests, gradient-boosted trees, and neural networks estimate parameters accurately without the distributional assumptions imposed by parametric methods, such as normally distributed errors in linear regression. However, the estimation process is often treated as a black box because the reasoning behind a fitted result is buried in thousands of trees or numerous neurons, which weakens the transparency and trustworthiness of what the model produces \parencite[]{khosravi2022explainable, lipton2018mythos,rudin2019stop}. For example, an unexplained predicted math achievement score is difficult to trust or to use for a parent weighing a course choice, a student deciding whether to persist, or a counselor allocating limited support. Learning analytics ultimately has to serve human sense-making, not merely optimize a metric.

Two obstacles compound this transparency gap. The first is interpretation. Even though the Explainable AI (XAI) community has proposed various post hoc explanation tools, their outputs, such as feature-importance bars, partial-dependence curves \parencite[]{friedman2001greedy}, and Shapley-value attributions \parencite[SHAP value;][]{lundberg2017unified}, are themselves specialized artifacts that presuppose statistical literacy. The second obstacle comes from model deployment. Running a tree ensemble or a neural network, maintaining its data preprocessing pipeline, and retrieving its post hoc explanations require hardware and expertise that classroom- and family-level stakeholders do not have \parencite[]{ifenthaler2017are, tsai2017learning}. Therefore, the opacity and operational burden could make a seemingly insightful learning analytics system the least able to engage with it.

Recent work suggests a different interface. Large language models (LLMs) have been demonstrated to act as predictive models for regression and classification tasks and can express their predictions in fluent natural language \parencite[]{dinh2022lift,hegselmann2023tabllm,song2025omnipred}, which raises the possibility of a single model that both predicts and narrates. Besides, post hoc interpretation tools like SHAP value attributions and functional ANOVA \parencite[fANOVA;][]{hooker2004discovering,hooker2007generalized} decompositions provide an individual-level account of any upstream model's output, and express it as a baseline (a grand mean) plus additive contributions from each covariate. A narration grounded in a prediction and its decomposition can tell a stakeholder not only the extent to which a student can benefit or be harmed by this educational intervention, but also which of that student's characteristics account for the difference, all in natural human language rather than in pure numbers.

Building on these, we propose a two-stage fine-tuning pipeline that distills a black-box ML estimator and its post hoc interpretations into a small, open-weight LLM that produces the estimate, the decisions, and the explanation in natural language. The fine-tuned LLM runs smoothly at inference time on local commodity hardware with no upstream dependencies. For example, a fine-tuned Gemma 4 model \parencite[]{team206gemma} at effective 2-billion parameter size (E2B) with 4-bit precision occupies roughly 3.6 GB on disk and peaks near 4.1 GB of memory during inference. Prior distillation work transfers a larger model's labels and free-form rationales \parencite[]{hinton2015distilling,hsieh2023distilling, ho2023large,magister2023teaching, shridhar2023distilling}, and evaluates the results in terms of task accuracy. What has not been done, to our knowledge, is to distill a fitted statistical estimator together with the decomposition that explains it, and then to audit the distilled model's narrations against that estimator rather than trusting that fluent text is accurate text.

Evaluating such a pipeline requires a ground truth against which to compare, which real data cannot provide. We therefore begin with a simulation study in which the true results are known by construction and cross-test upstream signals from both the oracle estimate and a realistic black-box estimator. The oracle condition establishes a ceiling and isolates what the distillation step loses from what the estimator loses. We then apply the pipeline to a causal ML task as a case study and estimate the effect of advanced mathematics coursework (Advanced Placement/International Baccalaureate; AP/IB) on four-year college enrollment. The pipeline itself is not specific to causal inference and accepts any fitted model that returns an individual-level estimate.

We make three contributions. First, we develop a faithfulness-first evaluation framework for self-explaining predictive LLMs, which audits every narration against the decomposition it claims to describe, checks whether the cited covariates are truly influential, and scores decisions against the accuracy of a trivial rule that always recommends the most common action. Second, the simulation study identifies where the proposed pipeline can be trusted. The key findings are that the distillation step itself is nearly lossless. What breaks trust is not the fine-tuned LLM but a noisy upstream estimator, whose errors the distilled model reproduces in confident and fluent prose. Third, we apply the proposed pipeline on a nationally representative dataset, the High School Longitudinal Study 2009 \parencite[HSLS:09;][]{ingels2011high}, and recover an established finding, that advanced mathematics coursework mostly benefits the students who are least likely to attend college \parencite[]{byun2015advanced}, and delivers it as audited, individualized explanations with a zero rate of fabricated statistics through the fine-tuned LLM.

The remainder of this paper proceeds as follows. The next section briefly reviews the three background areas: post hoc interpretation tools, LLM fine-tuning and distillation, and the estimation of heterogeneous treatment effects. The Method section introduces how we integrate all these methods into a proposed distillation pipeline. We then design a Monte Carlo simulation study to evaluate the estimation performance, fidelity, and faithfulness. The following empirical analysis demonstrates the pipeline's performance in a real-world setting. The last two sections talk about the advantages, lessons, limitations, and conclusions of the proposed method.

\section{Background}

\subsection{Interpretation tools} \label{sec:fANOVA}

A large family of post hoc methods has been developed to explain black-box predictions \parencite{guidotti2018survey, barredoarrieta2020explainable}, including local surrogate models \parencite{ribeiro2016should} and additive feature attributions such as SHAP \parencite{lundberg2017unified}. For understanding how a prediction surface depends on a covariate, the classical tool is the partial dependence plot, which averages predictions while a target covariate is set to fixed values \parencite{friedman2001greedy}, and its individual-level refinement, the individual conditional expectation curve \parencite{goldstein2015peeking}. Partial dependence, however, is only reliable when covariates are close to independent. When they are correlated, the averaging step forces the model to extrapolate into regions of the covariate space where little or no data exist, and the resulting curves can be badly misleading \parencite{hooker2021unrestricted, molnar2022general, molnar2023relating}. In educational datasets, this assumption is not always held since some commonly used variables, like student socioeconomic status, prior achievement, and school context, are rarely independent.

Accumulated local effects (ALE) were introduced to remove this failure mode \parencite{apley2020visualizing}. Instead of averaging predictions over the marginal distribution of the other covariates, ALE accumulates local finite differences computed within narrow conditional windows, so the model is only ever evaluated near observed data. The result is an estimate of each covariate's main-effect function that remains valid under correlated covariates, along with second-order functions that capture pairwise interactions. ALE therefore yields an additive decomposition of an estimate as a grand mean, plus a first-order term per covariate, plus higher-order remainder terms \parencite{apley2020visualizing, molnar2022interpretable}, which is easiest to state for one individual. Write the covariates as $x = (x_1, \ldots, x_p)$ and the fitted model's estimate as $\hat{f}(x)$. The retained terms then reconstruct the estimate as
\begin{equation}
\hat{f}(x) = g_0 + \sum_{j=1}^{p} g_j(x_j) + \sum_{j<k} g_{jk}(x_j, x_k) + \varepsilon(x),
\label{eq:fanova}
\end{equation}
where $g_0$ is the grand mean of the estimates across the sample, each first-order term $g_j$ is a centered curve that depends on one covariate alone, each second-order term $g_{jk}$ captures what a pair of covariates does beyond their separate curves, and the residual $\varepsilon(x)$ collects whatever the retained terms do not reproduce. ALE constructs each $g_j$ by accumulating the average local change in $\hat{f}$ across narrow windows of $x_j$, each average taken only over individuals actually observed in that window, and then centering the accumulated curve to mean zero \parencite{apley2020visualizing}. The second-order terms are built the same way from local changes in pairs of covariates. For illustration, one student's predicted probability of college enrollment might read $0.31 = 0.23 + 0.09 - 0.02 + 0.01$: a grand mean of $0.23$, plus $0.09$ from strong prior mathematics achievement, minus $0.02$ from low family income, plus a residual of $0.01$. Because the decomposition is computed from model outputs alone, it applies to any upstream learner, whether a tree ensemble, a neural network, or a stacked combination of several models. When the quantity being decomposed is itself a causal estimate, this decomposition acquires an additional substantive reading, which we discuss in Section~\ref{sec: hte}.

The individual-level decomposition also supports a variance-based importance measure. Functional ANOVA (fANOVA) splits the total variance of the estimate into the share carried by each covariate, the share carried by retained interactions, and a leftover \parencite[]{hooker2004discovering, sobol2001global}. Each ALE term is evaluated at every individual's observed values, which turns the term into a column of numbers across the sample, and the variance of that column becomes its share, so a covariate whose term barely moves across individuals receives a small share, however steep its curve looks alone. In symbols, the share assigned to covariate $j$ is
\begin{equation}
S_j = \frac{\widehat{\operatorname{Var}}\!\left[\,g_j(x_{ij})\,\right]}{\widehat{\operatorname{Var}}\!\left[\,\hat{f}(x_i)\,\right]},
\label{eq:share}
\end{equation}
with variances taken across individuals $i$ in the sample. Interaction shares $S_{jk}$ are defined the same way, and the remaining share reflects what is not captured by the retained terms, which absorbs both covariance cross-terms and residual variation. Because ALE components can be correlated with one another when input features are dependent \parencite[]{apley2020visualizing}, we read the shares $S_j$ as relative measures of how much of the additive account each term carries, rather than an exact partition of $Var(\hat{f})$. Variance-based importance of this kind is routine in applied machine learning \parencite[]{hutter2014efficient, watanabe2023pedanova}, and since we always compute shares on an ALE decomposition, we refer to the combination simply as ALE-fANOVA hereafter. The individual decomposition and the variance-based importance measure serve as the interpretive narrations that a fine-tuned LLM model needs to learn. 

SHAP values, as another common interpretative practice in the ML community, can also return a similar per-individual decomposition that sums to the prediction \parencite[]{lundberg2017unified}. Past studies demonstrated that the two tools agree in a pure additive setting, and that the algorithm used to estimate SHAP values is a transformation of the ALE-fANOVA decomposition \parencite[]{herren2022statistical}. However, a SHAP value splits interaction effects and absorbs them into individual variable scores  \parencite[]{bordt2023shapley}, and its no-leftovers rule requires the attributions to sum exactly to the estimate, so nothing is ever reported as unexplained. Our decomposition ledger needs those missing residuals, because the audit checks that the residual is declared rather than absorbed into confident narration. In addition, model-agnostic SHAP scales exponentially with the number of variables, and the accelerated TreeSHAP applies only to tree models \parencite[]{lundberg2020local}. We therefore rely on fANOVA as a primary interpretation tool and treat SHAP as a complement rather than a competitor.

\subsection{LoRA fine-tuning and distillation} \label{sec: lora}
LLM distillation compresses a capability from a larger mentor model into a smaller mentee model.\footnote{Computer science usually uses teacher--student to describe such a relation between models. To avoid confusion with the real teacher--student relationship in educational research, we use mentor--mentee.} Two developments make this practical on local hardware. Open-weight models are now released in sizes small enough to run on a commodity laptop, such as Google's Gemma \parencite[]{team206gemma} and Alibaba's Qwen \parencite[]{yang2025qwen3} families, and parameter-efficient fine-tuning trains a small set of added parameters while the pretrained network stays frozen \parencite[]{houlsby2019parameterefficient}. \textcite{hu2022lora} introduced low-rank adaptation (LoRA), which freezes the base weights and trains a pair of small adapter matrices for selected linear layers. Although LoRA updates well under 1\% of the model's parameters, studies demonstrate that it rivals full fine-tuning across model scales \parencite[]{ding2023parameterefficient, han2024parameterefficient, wang2025parameterefficient}.

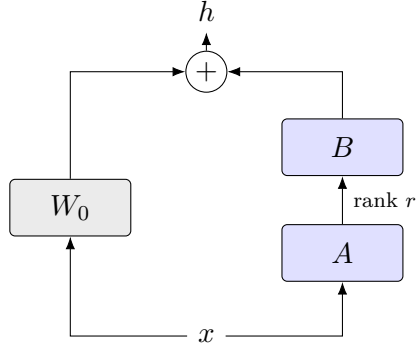
\begin{figure}[t]
\centering
\begin{tikzpicture}[font=\small, >=Latex,
  box/.style={draw, rounded corners=2pt, minimum width=1.6cm, minimum height=0.75cm, align=center},
  frozen/.style={box, fill=black!8}, train/.style={box, fill=blue!12}]
  \node (x) at (0,0) {$x$};
  \node[frozen] (w) at (-1.8,1.7) {$W_0$};
  \node[train]  (a) at (1.8,1.1)  {$A$};
  \node[train]  (b) at (1.8,2.5)  {$B$};
  \node[circle, draw, inner sep=1.5pt] (s) at (0,3.5) {$+$};
  \node (h) at (0,4.3) {$h$};
  \draw[->] (x) -- (-1.8,0) -- (w);
  \draw[->] (x) -- (1.8,0) -- (a);
  \draw[->] (a) -- (b) node[midway, right] {\scriptsize rank $r$};
  \draw[->] (w) -- (-1.8,3.5) -- (s);
  \draw[->] (b) -- (1.8,3.5) -- (s);
  \draw[->] (s) -- (h);
\end{tikzpicture}
\caption{LoRA at a single layer. The pretrained weights $W_0$ stay frozen (gray) while the low-rank adapters $A$ and $B$ (blue) are trained. The rank-$r$ bottleneck between them is what keeps the trainable parameter count small.}
\label{fig:lora}
\end{figure}

The mechanism is easiest to see at a single layer. The transformer, the core architecture behind LLMs, contains many weight matrices \parencite[]{vaswani2017attention}. Write one of them as $W_0 \in \mathbb{R}^{d \times k}$, mapping an input $x$ to an output $h$. Full fine-tuning would update all $dk$ entries of $W_0$. As shown in Figure~\ref{fig:lora}, LoRA leaves $W_0$ untouched and learns a low-rank correction alongside it, so the layer instead computes
\begin{equation}
h = W_0 x + s\,BAx, \qquad A \in \mathbb{R}^{r \times k},\quad B \in \mathbb{R}^{d \times r},\quad r \ll \min(d,k),
\label{eq:lora}
\end{equation}
where $r$ is a prespecified rank and $s$ a fixed scaling constant. Only the $r(d+k)$ entries of $A$ and $B$ are trained, rather than the $dk$ entries of $W_0$. For example, at $d = k = 2{,}048$ and $r = 16$, the adapter holds about $1.6\%$ as many parameters as the layer it modifies. Because $B$ is initialized at zero, $BA = 0$ at the start, so training begins from the base model exactly and the adapters accumulate only what the mentor traces\footnote{A trace is the written record of the reasoning steps by which a model reaches its output. Here, it is a scratchpad stating the arithmetic of the decomposition, the structured fields, and the prose that narrates them. We use the term in its reasoning-distillation sense rather than the learning analytics sense of learner behavioral logs. Appendix~\ref{app:traces} reproduces a complete mentor trace and the corresponding mentee output for one student.} teach.

To align the mentee model's output distribution with the mentor's, the training process minimizes the token-level cross-entropy of the mentee's output against the mentor trace,
\begin{equation}
\mathcal{L}(A, B) = -\sum_{t} \log p_{A,B}\!\left(y_t \mid y_{<t}, x\right),
\label{eq:loss}
\end{equation}
where $x$ is a student's covariate profile rendered as text and $y$ is the corresponding mentor trace. Gradients reach only $A$ and $B$. Since $W_0 + sBA$ has the same shape as $W_0$, the trained adapters can be folded into the base weights once the loss stabilizes, so deployment carries no extra parameters and no additional inference cost \parencite[]{hu2022lora}.

The closest existing designs treat the LLM as a translation layer for an upstream system. \textcite{zytek2024explingo} proposed a prompting pipeline that passes SHAP attributions to an LLM and grades the resulting narratives with another LLM, and recent work fine-tunes small open-weight models to narrate counterfactual explanations by distilling the narration skill from a larger LLM rather than from the fitted model itself \parencite[]{giorgi2025enhancing}. In both designs, the LLM restates an explanation another system provides and produces no estimate of its own. A recent audit of such prompted narrations in credit scoring documents found prompted narrations that flipped the signs of the attributions they were given, even with careful prompt design \parencite[]{pratama2026accurate}. We are therefore not aware of work that distills a fitted statistical estimator together with its decomposition-grounded explanation into a small open-weight model, and then audits the distilled model's narrations for arithmetic consistency, attribution fidelity, and decision fidelity against that estimator. The pipeline introduced in Section~\ref{sec:methods} fills that gap, and we instantiate it on a causal estimator as a case study.

\subsection{Estimating heterogeneous treatment effects} \label{sec: hte}

Learning analytics research has been urged to move beyond prediction toward causal questions \parencite[]{weidlich2022causal}, and we use causal ML as the case study for the proposed pipeline. Treatment heterogeneity matters for educational decision-making because intervention effects vary across students with different background profiles. Large-scale experiments show that the same intervention can produce substantial benefits for some subgroups and little effect for others \parencite[]{yeager2019national, kizilcec2020scaling,suk2026fair}, which a single average treatment effect (ATE) conceals. Researchers commonly summarize this heterogeneity with the conditional average treatment effect (CATE), the expected effect for the group of students who share a given set of characteristics.

Suppose we observe $n$ students with information on $(X, A, Y)$, where $X$ is a vector of student characteristics, $A$ is a binary treatment with $A = 1$ indicating the treated, and $Y$ is the observed outcome. Let $Y_i^{*}(a)$ denote the potential outcome that would be observed if student $i$ were assigned treatment $a \in \{0, 1\}$ \parencite[]{rubin1974estimating, imbens2015causal}. Because each student is observed under only one condition, the individual effect $Y_i^{*}(1) - Y_i^{*}(0)$ is never available, and we instead target
\begin{equation}
\tau(x) = \mathbb{E}\!\left[Y^{*}(1) - Y^{*}(0) \mid X = x\right],
\label{eq:cate}
\end{equation}
which is identified from observed data under consistency, conditional ignorability, and positivity \parencite[]{imbens2015causal, rubin1986comment}. We provide a detailed discussion on causal assumptions and identification in Appendix~\ref{app:identification}. A positive $\tau(x)$ indicates that students with profile $x$ benefit from the intervention on average.

Equation~\eqref{eq:cate} connects the case study back to the decomposition in Section~\ref{sec:fANOVA}. Averaging $\tau(x)$ over the covariate distribution returns the ATE, so the grand mean of a fitted CATE surface, i.e., the $g_0$ of Equation~\eqref{eq:fanova}, is a plug-in estimate of what the intervention does on average, and each first-order term describes how one moderator moves a particular student above or below that average. The decomposition therefore has a reading a stakeholder can follow without any statistical vocabulary, like here is the typical effect, and here are factors that drive students above or below it.

Many estimators are available for $\tau(x)$, including Bayesian additive regression trees and its causal extensions \parencite[]{chipman2010bart, hill2011bayesian, hahn2020bayesian}, causal forests \parencite[]{wager2018estimation, athey2019generalized}, and the metalearner family of S-, T-, X-, R-, and DR-learners \parencite[]{kunzel2019metalearners, nie2021quasioracle, kennedy2023optimal}. No single learner dominates across data regimes \parencite[]{knaus2021machine, curth2021nonparametric}. We use the X-learner for two reasons specific to our application. First, the X-learner was designed for imbalance treatment assignments and outperforms causal forests in the unbalanced designs of \textcite{kunzel2019metalearners}. Second, the X-learner is stable in estimation since it avoids division by near-zero propensity scores, a mechanism that often inflates estimation variance. We note that the choice of estimator is not essential to the pipeline, since any estimator that returns a per-student estimate can serve as the upstream model and the pipeline treats its output as a black box.

\section{Methods}\label{sec:methods}
\begin{figure}[htbp]
    \centering
    \includegraphics[width=1.0\textwidth]{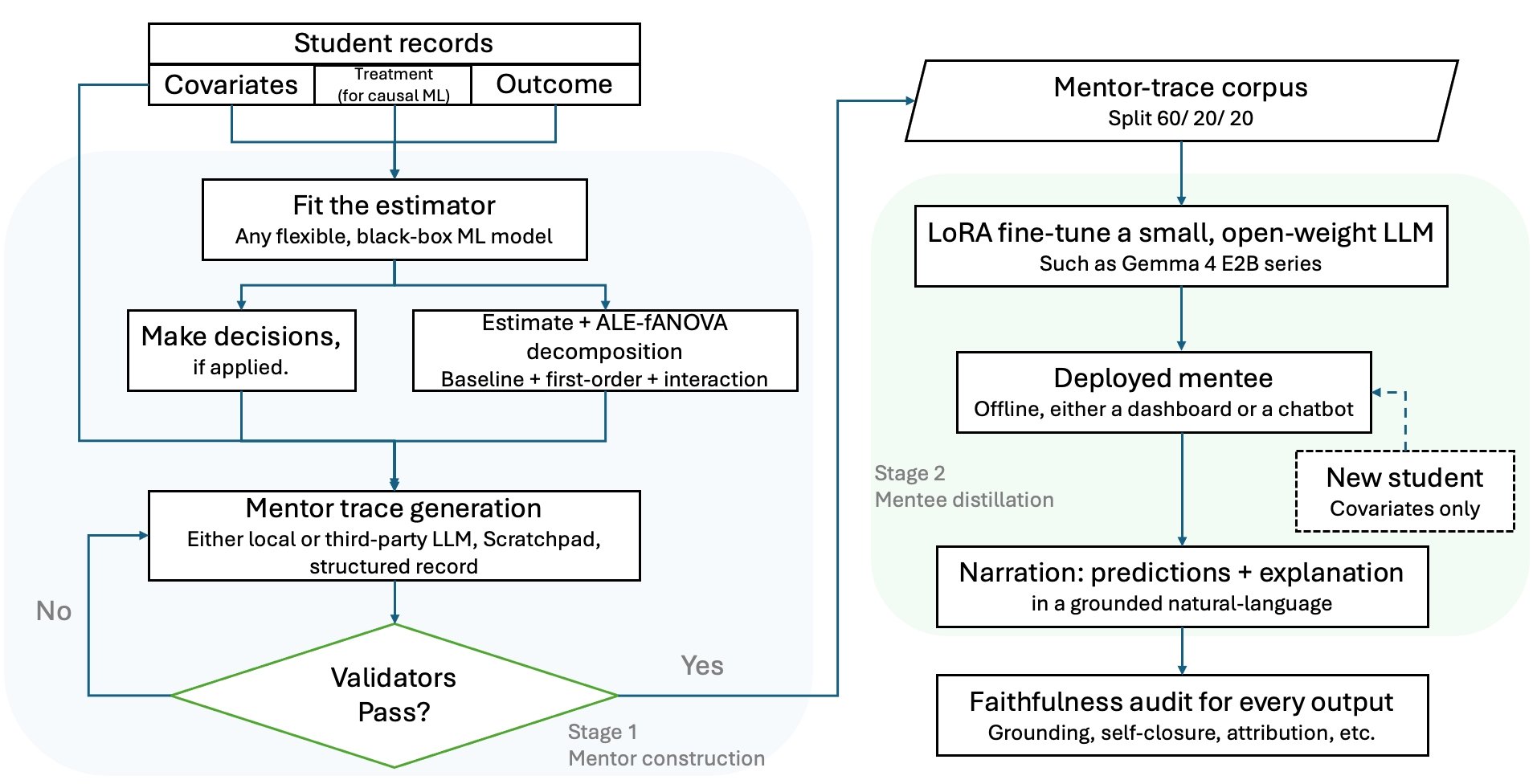}
    \caption{The two-stage distillation pipeline. In the mentor construction stage, any flexible ML model is fitted to the analytic sample to produce an individual-level estimate. ALE-fANOVA decomposes that estimate into a grand mean, first-order covariate terms, one data-driven pairwise interaction, and a residual. The decision-making procedure depends on the specific task. A capable mentor LLM renders these assets as a structured mentor trace, one per student, and the faithfulness audit (validators) acts as an acceptance gate so that only arithmetically closed, fully grounded traces enter the training corpus. In the mentee distillation stage, a small open-weight LLM is fine-tuned on the accepted corpus using low-rank adaptation (LoRA). The estimate, its decomposition, and the decision are absorbed into the mentee's weights, so that at deployment a single model maps a student's covariate profile to an effect estimate, an individualized explanation naming the characteristics responsible, and a recommended action, without the upstream estimator, the interpretation tool, or a network connection at inference time.}
    \label{fig:piepline}
\end{figure}

The proposed pipeline has two main stages, mentor construction and mentee distillation, as shown in Figure~\ref{fig:piepline}. The first stage fits a flexible ML model to get an estimate, interprets it using decomposition based on ALE-fANOVA, and derives a decision for each student. A capable mentor LLM translates all these assets into a structured output (i.e., a mentor trace). The second stage fine-tunes a small open-weight LLM on these traces with LoRA to produce a single model that maps a student's covariate profile to an effect estimate, an individualized explanation, and a recommended decision. The predictive and interpretation abilities are all distilled into the mentee's weights, so deployment requires only a small model that can be interacted with natural language rather than through a statistical interface.

\subsection{The mentor construction stage} \label{sec:mentor}

The first stage allows researchers and practitioners to use any preferred flexible ML model tailored to the specific research question, ranging from a neural network to a stacked approach that trains a group of candidate models with cross-validation \parencite[e.g., the Super Learner;][]{van2007super}. After obtaining the estimate for each individual, we apply the ALE-fANOVA approach and retain both first- and second-order terms, since interactions are commonly observed in educational settings. To construct the pairwise interaction, we bin the variance left unexplained by the first-order terms on the joint quantile grid of each pair of numeric covariates and retain the pair that accounts for the most residual variance. This data-driven screen finds the dominant interaction without requiring the analyst to specify it in advance. In the empirical application reported in Section~\ref{sec:empirical_analysis}, the first-order terms account for $67.7\%$ of the variance in the estimated effects, the retained interaction adds a further $6.6\%$, and the remaining $25.7\%$ is not captured by the retained terms, which might be the absorbing covariance and residual variation.
 
Each student's estimate is then reconstructed as the baseline (the grand mean), plus the cited first- and second-order contributions, plus a residual for whatever the retained terms do not capture. Panel A of Figure~\ref{fig:ledger} shows one example of the resulting decomposition ledger in the empirical sample. Two selection rules keep the citation set honest. A global eligibility filter admits a covariate as narratable only if its average absolute contribution reaches a fixed fraction of the strongest covariate's, which we set at $20\%$, and which prevents the narration from dignifying estimation noise with a name. Within a student, terms are then cited in descending order of absolute contribution until they account for $90\%$ of the total attribution, capped at five terms. 

\begin{figure}[htbp]
\centering
\caption{Decomposition ledger and generated Narration for one student from the mentor construction stage. This is part of the mentor trace that will ask a mentee model to learn from.}
\label{fig:ledger}
 
\footnotesize
\textbf{Panel A. Decomposition ledger}
 
\vspace{0.5em}
\begin{tabular}{llr}
\hline
Term & Observed value & Contribution \\
\hline
Baseline (grand mean) & --- & $+0.2289$ \\
School control & Catholic or other private & $-0.0463$ \\
Mathematics achievement & $70.41$ (above average) & $-0.0409$ \\
Socioeconomic status & $-0.373$ (below average) & $+0.0382$ \\
Student educational expectation & Complete a bachelor's degree & $+0.0306$ \\
Mathematics achievement $\times$ school climate & $70.41$, $-0.85$ & $-0.0156$ \\
Residual & --- & $+0.0201$ \\
\hline
Estimated effect & & $+0.2150$ \\
\hline
\end{tabular}
 
\vspace{1em}
\textbf{Panel B. Narration generated from Panel A}
 
\vspace{0.5em}
\begin{minipage}{0.92\textwidth}
\itshape
This student's predicted effect of $+0.2150$ is slightly below the average effect of $+0.2289$. The strongest downward driver is school control, observed as ``Catholic or other private,'' which lowers the benefit by $-0.0463$; the mathematics standardized theta score, observed at $70.41$ (above average), further lowers the benefit by $-0.0409$, in line with the model's general pattern that lower mathematics standardized theta score is associated with larger benefits. Conversely, the socio-economic status composite, observed at
$-0.373$ (below average), raises the benefit by $+0.0382$, in line with the model's general pattern that lower socio-economic status composite is associated with larger benefits, and the student's expectation ``Complete a Bachelor's degree'' adds $+0.0306$. The interaction between mathematics standardized theta score ($70.41$, above average) and school climate scale ($-0.85$, below average) dampens the effect by $-0.0156$, in line with the model's finding of a joint effect beyond their separate contributions. Because the effect is clearly distinguishable from zero, the recommendation is to enroll the student in AP/IB math coursework.
\end{minipage}
 
\vspace{0.75em}
\begin{minipage}{0.92\textwidth}
\raggedright
\textit{Note.} Contributions are on the probability scale of four-year college enrollment. The ledger closes exactly: $0.2289 - 0.0463 - 0.0409 + 0.0382 + 0.0306 - 0.0156 + 0.0201 = 0.2150$. School control and student educational expectation are categorical and therefore carry no interpretive clause in Panel B (see Table~\ref{tab:pool}). The mentee is trained on the narration in Panel B, not on the ledger in Panel A.
\end{minipage}
\end{figure}

The final step of Stage~1 turns each student's ledger into the narration the mentee will learn from. A capable mentor LLM writes flowing prose that situates the student's estimate against the grand mean and then, for every cited term, states the student's observed value on its original scale together with the signed contribution that value receives in the decomposition, closing with the recommended decision in actionable terms\footnote{Narrations can be generated by any capable LLM, hosted or local. We used DeepSeek-V4-Flash for its low cost and high request concurrency. Generating roughly $7{,}300$ accepted narrations cost under \$4. A locally served open-weight model can be substituted when data privacy requires it.}. The only interpretive language the mentor may add comes from a small phrase pool derived from the fitted model, as illustrated in Table~\ref{tab:pool}. For each narratable numeric covariate, we summarize the direction of its association with the estimate across the sample and store a single sentence expressing that tendency. We also provide a short description of the retained interaction. Therefore, the mentor cannot explain a contribution by hallucinating a mechanism. Rather, it can only restate the attribution pattern the estimator actually learned. Panel B of Figure~\ref{fig:ledger} gives the narration generated from the ledger in Panel A, in which every number and every interpretive clause traces to Panel A or to Table~\ref{tab:pool}. Each narration also carries a scratchpad enforcing arithmetic closure, $\text{baseline} + \text{cited contributions} + \text{residual} = \text{estimate}$. An LLM will produce fluent narration even if it is false, so we design validators to audit faithfulness as discussed in Section~\ref{sec: audit_gate}. 


\begin{table}[htbp]
\centering
\caption{Phrase Pool Derived From the Fitted Estimator}
\label{tab:pool}
\footnotesize
\begin{tabular}{p{0.30\textwidth} c p{0.46\textwidth}}
\hline
Covariate & Direction & Phrase available to the mentor \\
\hline
Mathematics achievement & $-$ &
lower mathematics achievement is associated with larger benefits \\
Socioeconomic status & $-$ &
lower socioeconomic status is associated with larger benefits \\
School climate (administrator-assessed) & $+$ &
higher school climate is associated with larger benefits \\
School staff expectations (counselor-reported) & $-$ &
lower staff expectations are associated with larger benefits \\
Percent of 12th graders in AP courses & $-$ &
a lower percentage of 12th graders in AP courses is associated with larger benefits \\
Mathematics achievement $\times$ school climate & n/a &
a joint effect between mathematics achievement and school climate beyond their separate contributions \\
\hline
\end{tabular}
 
\vspace{0.5em}
\begin{minipage}{0.95\textwidth}
\raggedright
\footnotesize
\textit{Note.} Direction is the sign of the covariate's association with the
estimated effect across the sample. Each phrase is generated from the fitted
estimator, not written by the mentor, and the mentor may attach a phrase only
to a term the ledger already cites. Five of the ten eligible covariates are
categorical and receive no phrase.
\end{minipage}
\end{table}

\subsection{The mentee distillation stage}

We adapt an open-weight Gemma 4's E2B variant with the LoRA approach implemented in \texttt{mlx-lm} (v0.31.3) on a MacBook Pro laptop with an M5 Max chip and 128GB unified memory. The Gemma E2B model has about 2 billion active parameters. At 16-bit precision, it occupies 10.3~GB of disk space, 10.7~GB of peak memory usage during inference, and 24.9~GB during fine-tuning. All base weights are frozen, and rank-16 low-rank adapters are injected into the linear projection layers of the top 16 transformer blocks. With a scaling factor $s=20$ and an adapter dropout of 0.05, we have about 13.6 million trainable parameters, which is only about 0.7\% of the base model. 

Each training example includes a fixed system prompt that establishes the causal-inference analyst role, a user prompt containing only the subject's covariates rendered as a structured feature block, and an assistant target equal to the mentor trace, which serves as a supervisory outcome. See Appendix~\ref{app:traces} for a comprehensive example. We optimize the token-level cross-entropy by computing it only over the assistant tokens and masking out the system and user prompt tokens, so that the Gemma model is supervised to generate the trace rather than to reproduce the prompt. We use AdamW with a peak learning rate of $5\times10^{-5}$, a cosine-decay schedule and a 100-step warmup, a batch size of 2, a maximum sequence length of 3{,}200 tokens, and 3{,}000 iterations with a fixed seed, keeping the final checkpoint. In preliminary runs, selecting among intermediate checkpoints by validation loss never beat the final checkpoint by a meaningful margin, so the final design drops checkpoint screening. One full fine-tune takes roughly 35 minutes on the laptop described above.

\subsection{Faithfulness audit} \label{sec: audit_gate}

An LLM will produce a fluent narration whether or not it is true, so we measure faithfulness across four layers, and the same faithfulness audit procedure serves in both stages. At corpus-construction time, it serves as the acceptance gate for mentor traces and repairs questionable outputs until they pass all layers, while at evaluation time it applies to everything the mentee generates, without a repair step. The first layer checks that the stated decomposition must close arithmetically (i.e., baseline plus cited contributions plus residual equals the stated estimate), every cited feature must appear in the prose with its observed value and its signed contribution, and every number in the prose must be traceable to the decomposition. The second layer ensures that the cited covariates are the influential ones from the upstream, rather than hallucination. In simulation, where the true influential variables are known, we score this with the area under the ROC curve (AUC) for ranking them above noise, and the rate at which noise and true interactions are cited. In the empirical application, where ground truth is unavailable, we measure agreement with the mentor's attributions instead. The third layer focuses on decision fidelity by comparing the mentee's recommended action with the action implied by the ground truth (in simulation) or by the mentor (empirically). The last layer pays particular attention to the most consequential failure, which recommends treatment for individuals the intervention would actually harm.

\section{Simulation Study}

A simulation study is essential to evaluate the faithfulness of the proposed pipeline in the causal ML case, as individual treatment effects are never observed in empirical data, whereas simulations can provide the ground truth by construction. In this section, we focus on two primary questions: (1) With the oracle signal, will the fine-tuning process faithfully recover the ground truth, including the estimate, the critical covariates, and the decompositions? (2) Under the given estimator, how will the performance change? 

\subsection{Data generating process}
To match the number of covariates in the empirical data, we generate 21 covariates following a causal structure, in which three true moderators $\{X_1, X_2, X_3\}$, three confounders $\{X_1, X_2, X_5\}$, one outcome predictor $X_4$, and 16 covariates serve as pure noise. $X_4$ and $X_5$ serve as two decoys to test whether the pipelines can distinguish them from the true moderators. All covariates are independently and identically sampled from standard normal distributions, except $X_5$, which is a binary variable generated from a Bernoulli distribution with a probability of 0.5, and $X_6$, which is a categorical variable with four categories uniformly generated. Table~\ref{tab:study2_sim_covaraites} provides a summary of the covariates and their roles. 

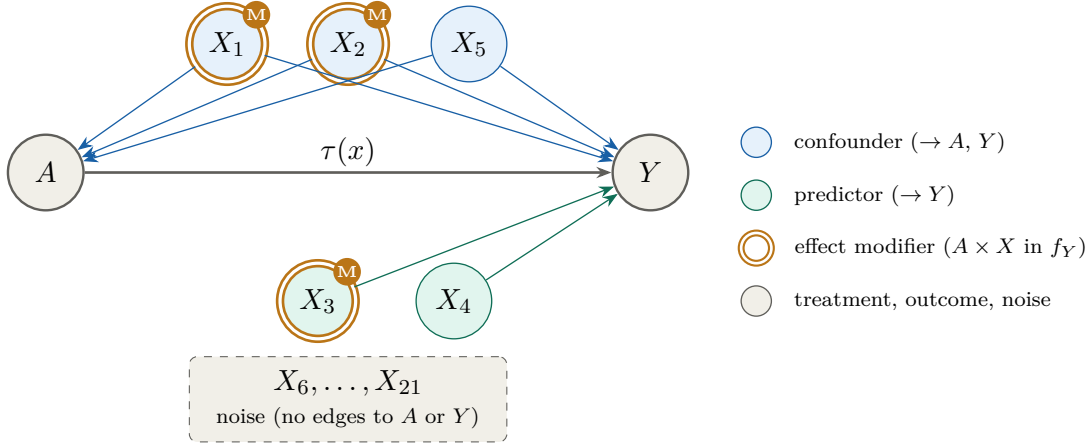
\begin{figure}[ht]
\centering
\begin{tikzpicture}
 
  \node[focal] (A)  at (0,1.7)   {$A$};
  \node[focal] (Y)  at (8,1.7)   {$Y$};
 
  \node[conf, mod] (X1) at (2.4,3.4) {$X_1$};
  \node[conf, mod] (X2) at (4.0,3.4) {$X_2$};
  \node[conf]      (X5) at (5.6,3.4) {$X_5$};
 
  \node[pred, mod] (X3) at (3.6,0) {$X_3$};
  \node[pred]      (X4) at (5.4,0) {$X_4$};
 
  \node[noise, minimum width=4.2cm] (N) at (4,-1.3)
        {$X_6,\dots,X_{21}$\\[-1pt]{\scriptsize noise (no edges to $A$ or $Y$)}};
 
  \draw[cedge] (X1) -- (A);  \draw[cedge] (X1) -- (Y);
  \draw[cedge] (X2) -- (A);  \draw[cedge] (X2) -- (Y);
  \draw[cedge] (X5) -- (A);  \draw[cedge] (X5) -- (Y);
  \draw[pedge] (X3) -- (Y);
  \draw[pedge] (X4) -- (Y);
  \draw[tedge] (A) -- node[above, font=\small, inner sep=2pt] {$\tau(x)$} (Y);
 
  \node[lkey, fill=confcol, draw=confdraw] (k1) at (9.4,2.1) {};
  \node[right=1.5mm of k1, font=\scriptsize, anchor=west] {confounder ($\to A,\,Y$)};
  \node[lkey, fill=predcol, draw=preddraw] (k2) at (9.4,1.4) {};
  \node[right=1.5mm of k2, font=\scriptsize, anchor=west] {predictor ($\to Y$)};
  \node[lkey, draw=modcol, line width=0.8pt, double, double distance=1pt] (k3) at (9.4,0.7) {};
  \node[right=1.5mm of k3, font=\scriptsize, anchor=west] {effect modifier ($A\times X$ in $f_Y$)};
  \node[lkey, fill=neutcol, draw=neutdraw] (k4) at (9.4,0.0) {};
  \node[right=1.5mm of k4, font=\scriptsize, anchor=west] {treatment, outcome, noise};
 
\end{tikzpicture}
\caption{Causal graph for the simulation design. Edges encode direct causal
effects only; each variable appears once. Effect modification is \emph{not} a
graph edge---it is a property of the structural equation $f_Y$, marked here by
the amber ring. The moderators are exactly the covariates entering $\tau(x)$,
namely $\{X_1,X_2,X_3\}$: $X_1,X_2$ are confounders that also modify the effect,
$X_3$ is a non-confounding predictor that modifies it, while $X_5$ (confounder)
and $X_4$ (predictor) do not. $X_6,\dots,X_{21}$ are independent noise.}
\label{fig:dgp}
\end{figure}

\begin{table}[!h]
\caption{Specification of covariates in the data-generating process.}
\label{tab:study2_sim_covaraites}
\centering
\footnotesize
\begin{tabular}{c c c}
\hline
Label & Distribution & Role \\
\hline
$X_1$ & $\mathcal{N}(0,1)$ & moderator / confounder \\
$X_2$ & $\mathcal{N}(0,1)$ & moderator / confounder \\
$X_3$ & $\mathcal{N}(0,1)$ & moderator \\
$X_4$ & $\mathcal{N}(0,1)$ & outcome predictor \\
$X_5$ & $\mathrm{Bernoulli}(0.5)$ & confounder \\
$X_6$ & $\mathrm{Uniform}\{0,1,2,3\}$, one-hot & pure noise (categorical) \\
$X_7 \dots X_{21}$ & $\mathcal{N}(0,1)$ & pure noise \\
\hline
\end{tabular}
\end{table}

The overall outcome model is $$Y = \mu_0(X)+A\cdot\tau(X)+\epsilon, \epsilon\sim\mathcal{N}(0, 0.3^2)$$ where the baseline effect is $$\mu_0(x) = 0.5x_1+0.4x_4+0.3x_2+0.2x_5.$$  To precisely control for the variance shares in the true treatment effect model $\tau(x)$, we first define two base functions as $L(X_1)=\sigma(2X_1)-0.5$ and $D(X_2)=-X_2$ and normalize each by its own standard deviation to get $\tilde{L}(X_1)$ and $\tilde{D}(X_2)$ with unit-variance. We set $$\tau(x) = \tilde{L}(X_1) + \tilde{D}(X_2) + \gamma(X_1 \cdot X_3)$$ with coefficient $\gamma^2 =2$, so the interaction term carries variance 2 against the unit-variance main effects, which gives exactly 50\% variance share to it and 25\% to each of $X_1$ and $X_2$. The decision rule is to treat if $\tau(x)>0$. These values will serve as ground-truth criteria for comparison with the pipeline's results. The propensity model is $$e(x)=\sigma(0.6x_1+0.4x_2+0.3x_5),$$ where $\sigma(z) = \frac{1}{1+e^{-z}}$ is the standard sigmoid function.

We intentionally manipulate the true decision distributions to construct a moderately imbalanced case with 72.5\% treated observations and a severely imbalanced case with 97.1\% treated observations since the imbalanced distributions are not uncommon in educational data. We set the sample size to $n=9,167$ to be consistent with the empirical dataset and run for five rounds of repetitions.

\subsection{Evaluation Criteria}\label{sec:sim_eval}

We evaluate the pipeline on a held-out test set ${(x_i,\tau_i)}_{i=1}^{n}$, where $\tau_i \equiv \tau(x_i)$ is the known ground-truth CATE of individual $i$. Two mentor conditions are compared. The oracle mentor provides the true effect surface $\tau(\cdot)$ directly and thus serves as the ceiling performance, which isolates the fidelity of the distillation step from estimation error. The realistic mentor is an X-learner \parencite[]{kunzel2019metalearners}, chosen for its stable CATE estimation. For each test individual $i$ in both conditions, the mentee produces a point estimate $\hat{\tau}_i$, a set $\mathcal{C}_i$ of cited terms, and binary decisions. Every metric compares these outputs against the ground truth $\tau_i$, never against the mentor, so that fidelity is assessed at the level of the true effect. Therefore, the simulation study is a 2-by-2 design of $\{\text{severely imbalanced, moderately imbalanced}\} \times \{\text{oracle CATE, X-learner CATE}\}$ .

Let $\mathcal{M}=\{1,2,3\}$ index the true moderators $\{X_1,X_2,X_3\}$ and let $\mathcal{M}^{c}$ collect the two decoys $\{X_4,X_5\}$, which are correlated with treatment or outcome and so most easily mistaken for moderators, together with the noise covariates $\{X_6,\dots,X_{21}\}$. Our metrics fall into four dimensions: (1) point estimation fidelity, (2) attribution accuracy, (3) decision accuracy and safety, and (4) narration faithfulness. In the first dimension, we summarize how well the mentee's predicted effects track the truth with the Pearson correlation between $\{\hat{\tau}_i\}$ and $\{\tau_i\}$,
\begin{equation}
r \;=\;
\frac{\sum_{i=1}^{n}\bigl(\hat{\tau}_i-\bar{\hat{\tau}}\bigr)\bigl(\tau_i-\bar{\tau}\bigr)}
{\sqrt{\sum_{i=1}^{n}\bigl(\hat{\tau}_i-\bar{\hat{\tau}}\bigr)^{2}}\;
 \sqrt{\sum_{i=1}^{n}\bigl(\tau_i-\bar{\tau}\bigr)^{2}}},
\end{equation}
where $\bar{\hat{\tau}}$ and $\bar{\tau}$ are sample means. Values near $1$ indicate that the ordering and relative magnitude of individual effects are recovered. However, a correlation can be high even when every estimate is pulled toward the sample mean. Therefore, we report the slope of the least-squares fit of $\hat{\tau}$ on $\tau$. A slope of $1$ means magnitudes are preserved, while a slope well below $1$ means the mentee systematically compresses large effects and inflates small ones, which is a regression-to-the-mean issue we call magnitude compression in this study. We also provide the root mean squared error (RMSE) to compare the estimation accuracy across settings.

The second dimension focuses on attribution accuracy. We use two metrics to assess whether the explanation attributes the effect to the correct covariates. Let $s_j\ge 0$ denote the importance the pipeline assigns to covariate $j$, taken as the magnitude of its attributed contribution aggregated over the test set. The moderator-discrimination AUC is the probability that a true moderator outranks a non-moderator,
\begin{equation}
\mathrm{AUC} \;=\;
\frac{1}{|\mathcal{M}|\,|\mathcal{M}^{c}|}
\sum_{j\in\mathcal{M}}\;\sum_{k\in\mathcal{M}^{c}}
\Bigl[\mathbbm{1}(s_j>s_k)+\tfrac{1}{2}\,\mathbbm{1}(s_j=s_k)\Bigr],
\end{equation}
where $\text{AUC} = 1$ indicates that all three moderators are ranked above every decoy and noise covariate, and $\text{AUC} = 0.5$ corresponds to random guessing. Because the AUC is a ranking summary that can mask spurious attributions in lower-ranked slots, we pair it with a decoy/noise citation rate to measure the average proportion of mentee's cited covariates that are not true moderators,
\begin{equation}
\mathrm{FalseCite} \;=\;
\frac{1}{n}\sum_{i=1}^{n}\frac{\bigl|\mathcal{C}_i\cap\mathcal{M}^{c}\bigr|}{|\mathcal{C}_i|}.
\end{equation}
AUC and FalseCite ask whether the right covariates are cited. We further define closure metrics to measure how much of the effect the decomposition behind those citations can represent. Our narration explains each estimate as a sum of a baseline, the retained ALE terms, and a residual, $\hat{\tau}_i = \mu + \sum_{g\in\mathcal{C}_i} g(x_i) + \varepsilon_i$, where $\mu$ is the average effect (i.e., the grand mean) and the retained set contains the first-order terms together with the interaction selected by the screen discussed in Section~Methods. For any set $\mathcal{R}$ of retained terms, define the closure of the corresponding reconstruction $\tilde{f}_{\mathcal{R}}(x)=\mu+\sum_{g\in\mathcal{R}} g(x)$ against the surface $f$ it decomposes as
\begin{equation} \label{eq:closure}
\rho^{2}_{\mathcal{R}}
\;=\;
\operatorname{corr}^{2}\!\Bigl(\{\tilde{f}_{\mathcal{R}}(x_i)\}_{i=1}^{n},\,\{f(x_i)\}_{i=1}^{n}\Bigr),
\end{equation}
where $f=\tau$ in the oracle condition and $f=\hat{\tau}$ for the X-learner. We work on the squared (variance) scale so that closure values read as shares of effect variance, and we always evaluate the full retained decomposition rather than only the terms cited for a particular student, which separates genuine non-additivity from citation truncation. We define Equation~\ref{eq:closure} into two special cases. The first-order closure $\rho^{2}_{\mathrm{add}}$ retains only the additive terms. It measures how much interaction structure a surface contains and serves as the baseline against which the interaction screen ranks candidate pairs. A value near $1$ indicates that an additive narration can speak of all variance in the effect. The decomposition closure $\rho^{2}_{\mathrm{dec}}$ retains the additive terms plus the selected interaction. Because these are exactly the ingredients the narration can cite, the decomposition closure represents the upper-bound of the fidelity attainable by the explanation, with any shortfall absorbed into the residual $\varepsilon_i$. The gap between the two closures is precisely the share of the effect variance attributable to the detected interaction. A low first-order closure $\rho^{2}_{\mathrm{add}}$ paired with a high decomposition closure $\rho^{2}_{\mathrm{dec}}$ indicates the interaction screen found the structure that matters, while low values of both would suggest the surface contains structure the narration cannot represent and that the explanation should not be trusted to carry it.

The third dimension is decision quality and safety. The mentee's decision $\hat{d}_i$ follows the study's decision rule applied to its own estimate, $\hat{d}_i = \mathbbm{1}(\hat{\tau}(x) > 0) $, and the oracle decision $d_i$ applies the same rule to the known truth $\tau(x)$. Decision accuracy is the agreement between the two,
\begin{equation}
\mathrm{DecAcc} \;=\; \frac{1}{n}\sum_{i=1}^{n}\mathbbm{1}\bigl(\hat{d}_i=d_i\bigr).
\end{equation}
Because not all errors are equally costly, we define the unsafe-treatment rate as the proportion of individuals wrongly recommended for treatment among those whom the treatment actually harms, 
\begin{equation}
\mathrm{UnsafeTreat} \;=\;
\frac{\sum_{i=1}^{n}\mathbbm{1}\bigl(\hat{d}_i=1\;\text{and}\;\tau_i<-\delta\bigr)}
{\sum_{i=1}^{n}\mathbbm{1}(\tau_i<-\delta)},
\end{equation}
where $\delta$ is a customizable threshold, set to $0$ by default and to $0.05$ on the scale of absolute difference in probability (i.e., risk difference) to better handle imbalanced cases. This UnsafeTreat isolates the most consequential failure mode for educational decision support, and a trustworthy pipeline should hold it at or near zero.

The last dimension measures the faithfulness of the mentee's narration from three aspects. The Grounding Audit is about the proportions of narrations in which every number and category matches the model's own produced decomposition, with no fabricated values. Self-closure is the proportion about whose stated terms sum to the stated effect, similar to the arithmetic closure shown in panel~A in Figure~\ref{fig:ledger}. Uniqueness is about the proportion of distinct narrations across students.

\subsection{Results}\label{sec:sim_results}

Table~\ref{tab:sim_richer} reports the 2-by-2 design crossing the severely or moderately imbalanced true decision distributions with the upstream signal from the oracle CATE or the X-learner's estimated CATE. Every metric is computed against the ground-truth CATE and is reported for both mentee and mentor, from which we can compare the estimator loss to the distillation loss.

\begin{table}[!h]
\caption{Simulation results (mean$\pm$SD over $r{=}5$ independent draws) under the severely imbalance case (97.1\% are truly treated ) and moderately imbalanced case (72.5\% are truly treated). Every metric is computed against the ground-truth CATE, and each is reported for the mentee and for the mentor that taught it, so that estimator loss (mentor vs. truth) is separated from distillation loss (mentee vs. mentor). Decisions rule is to recommend treatment if $\hat\tau>0$. Harmful and beneficial subgroups use the margin $\delta=0.05$. The majority baseline is the accuracy of always recommending the majority action. Rows marked $\dagger$ are fixed by construction in the oracle columns, where the mentor is the ground truth. Rows marked $\ddagger$ are properties of the mentor's decomposition and have no mentee counterpart. The mentor's narration rows are $1.000$ by construction, because the validators act as an acceptance gate and only traces that pass enter the corpus.}
\label{tab:sim_richer}
\centering
\footnotesize
\begin{tabular}{l cc cc}
\hline
 & \multicolumn{2}{c}{Imbalanced case} & \multicolumn{2}{c}{Balanced case}\\
\cmidrule(lr){2-3}\cmidrule(lr){4-5}
Metric & Oracle & X-learner & Oracle & X-learner\\
\hline
\multicolumn{5}{l}{\emph{Point fidelity}}\\
Pearson $r$ (mentee)   & $0.925\pm.064$ & $0.723\pm.038$ & $0.910\pm.061$ & $0.729\pm.045$\\
Pearson $r$ (mentor)$^\dagger$ & $1.000\pm.000$ & $0.654\pm.027$ & $1.000\pm.000$ & $0.659\pm.028$\\
Slope (mentee)         & $0.902\pm.040$ & $0.495\pm.056$ & $0.918\pm.028$ & $0.498\pm.053$\\
Slope (mentor)$^\dagger$ & $1.000\pm.000$ & $0.535\pm.056$ & $1.000\pm.000$ & $0.539\pm.057$\\
RMSE (mentee)          & $0.036\pm.015$ & $0.070\pm.005$ & $0.041\pm.016$ & $0.070\pm.006$\\
RMSE (mentor)$^\dagger$ & $0.000\pm.000$ & $0.078\pm.003$ & $0.000\pm.000$ & $0.078\pm.003$\\
\hline
\multicolumn{5}{l}{\emph{Attribution}}\\
Moderator AUC (mentee) & $1.000\pm.000$ & $0.989\pm.017$ & $1.000\pm.000$ & $0.993\pm.010$\\
Moderator AUC (mentor) & $1.000\pm.000$ & $0.996\pm.008$ & $1.000\pm.000$ & $0.996\pm.008$\\
FalseCite (mentee)     & $0.000\pm.000$ & $0.263\pm.240$ & $0.000\pm.000$ & $0.263\pm.242$\\
FalseCite (mentor)     & $0.000\pm.000$ & $0.251\pm.229$ & $0.000\pm.000$ & $0.249\pm.228$\\
Interaction recovery (mentee) & $0.786\pm.015$ & $0.599\pm.145$ & $0.789\pm.020$ & $0.634\pm.214$\\
Interaction recovery (mentor) & $0.804\pm.017$ & $0.626\pm.139$ & $0.804\pm.017$ & $0.644\pm.137$\\
First-order closure $\rho^{2}_{\mathrm{add}}$$^\ddagger$ & $0.496\pm.012$ & $0.729\pm.029$ & $0.496\pm.012$ & $0.728\pm.029$\\
Decomposition closure $\rho^{2}_{\mathrm{dec}}$$^\ddagger$ & $0.969\pm.002$ & $0.787\pm.019$ & $0.969\pm.002$ & $0.789\pm.018$\\
\hline
\multicolumn{5}{l}{\emph{Decision and safety}}\\
DecAcc (mentee)        & $0.982\pm.006$ & $0.971\pm.004$ & $0.942\pm.014$ & $0.824\pm.041$\\
DecAcc (mentor)$^\dagger$ & $1.000\pm.000$ & $0.966\pm.006$ & $1.000\pm.000$ & $0.801\pm.019$\\
Majority baseline      & $0.971\pm.004$ & $0.971\pm.004$ & $0.725\pm.010$ & $0.725\pm.011$\\
Decision AUC (mentee)  & $0.970\pm.026$ & $0.823\pm.067$ & $0.993\pm.007$ & $0.939\pm.020$\\
Decision AUC (mentor)$^\dagger$ & $1.000\pm.000$ & $0.808\pm.066$ & $1.000\pm.000$ & $0.909\pm.025$\\
UnsafeTreat (mentee)   & $0.289\pm.087$ & $1.000\pm.000$ & $0.028\pm.020$ & $0.365\pm.144$\\
UnsafeTreat (mentor)$^\dagger$ & $0.000\pm.000$ & $0.926\pm.052$ & $0.000\pm.000$ & $0.386\pm.118$\\
\hline
\multicolumn{5}{l}{\emph{Narration faithfulness}}\\
Grounding audit (mentee) & $0.998\pm.002$ & $0.998\pm.001$ & $0.998\pm.003$ & $0.998\pm.001$\\
Grounding audit (mentor) & $1.000$ & $1.000$ & $1.000$ & $1.000$\\
Self-closure (mentee)  & $0.997\pm.003$ & $0.999\pm.001$ & $0.995\pm.005$ & $0.999\pm.001$\\
Self-closure (mentor)  & $1.000$ & $1.000$ & $1.000$ & $1.000$\\
Uniqueness (mentee)    & $1.000\pm.000$ & $1.000\pm.000$ & $1.000\pm.000$ & $1.000\pm.000$\\
Uniqueness (mentor)    & $1.000\pm.000$ & $1.000\pm.000$ & $1.000\pm.000$ & $1.000\pm.000$\\
\hline
\end{tabular}
\end{table}

Under the oracle mentor, the mentee recovers the true CATE surface closely. Pearson correlations reach $0.925$ and $0.910$, and slopes reach $0.902$ and $0.918$ in the two decision distributions, so the E2B Gemma model fine-tuned on roughly seven thousand narrations both ranks students correctly and reproduces the scale of the effects rather than collapsing them toward a constant. The pipeline recovers the decomposition structure and critical variables as well. First-order terms alone close $\rho^{2}_{\mathrm{add}} = 0.496$ of the variance in the true CATE, and adding the narrated interaction raises decomposition closure to $\rho^{2}_{\mathrm{dec}} = 0.969$, which confirms that the second-order term carries roughly half of the structure the DGP is built to contain. The mentee cites that interaction ($X_1\times X_3$) for about $79\%$ of test students against the mentor's $80\%$, ranks the true moderators above the decoys and the noise covariates at $\text{AUC} = 1.000$, and never cites a covariate from the decoys and noises ($\text{FalseCite} = 0.000$). With a clean upstream signal, the pipeline learns both the prediction and the post hoc interpretation of it.

After switching the estimator to X-learner, the pipeline performance change comes almost entirely from the upstream. The mentee's correlation with the truth falls to about $0.72$, and its slope drops to about $0.50$, which reflects a compression toward the mean due to the finite-sample estimation limitation of the X-learner. Decomposition and critical variable detection degrade as well. Interaction recovery drops to $0.599$ and $0.634$, and FalseCite rises to $0.263$, so each student-level decomposition cites at least one decoy or noise variable on average. Moderator ranking is the exception, holding at $\text{AUC} \geq 0.989$, which indicates the mentee still separates real moderators from decoys and noise even when it misstates their magnitudes. The performance degradation can be clearly explained by the metrics of the paired mentor, whose own slope against the truth is $0.535$ and $0.539$, so the compression is present in the estimator before any narration is written, and every mentee--mentor gap in these two columns is at most $0.074$. The bottleneck comes from the upstream estimator, not the fine-tuning. We remark that the mentee's point estimation is closer to the ground truth than the mentor's, with correlations of $0.723$ and $0.729$ against the mentor's $0.654$ and $0.659$, and a lower RMSE in both regimes, which are caused by the ALE decomposition process and the LLM distillation process that smooth the noisy CATE.

The metrics under the decision and safety dimensions reveal that the mentee model inherits the mentor's performance in decision making, as the gaps in $\text{DecACC}$ and $\text{Decision AUC}$ are lower than $0.06$. However, under the severely imbalanced case with the X-learner, the mentee recommends treatment for every such student ($\text{UnsafeTreat} = 1.000$) where the mentor already recommends treatment for $0.926$ of them, so the estimator introduces most of the failure and distillation closes the remainder. Under the oracle mentor, the mentor is correct by construction and the mentee still misrecommends $28.9\%$ of the harmed subgroup, which is pure distillation loss specific to the severely imbalanced case, and it is the largest such gap anywhere in the table. In the balanced case, where the harmed subgroup is larger and less exceptional, the pipeline demonstrates lower distillation loss at $2.8\%$ under the oracle CATE, and most of the unsafe recommendations inherit from the estimator X-learner, as the $\text{UnsafeTreat}$s are 0.365 and 0.386 for mentee and mentor, respectively.

The metrics under narration faithfulness are close to $1$, which shows that the mentee learns to write arithmetically closed decompositions and that the non-hallucinated narrations are essentially preserved.

\section{Empirical Analysis} \label{sec:empirical_analysis}
We apply the proposed pipeline to the High School Longitudinal Study of 2009 \parencite[HSLS:09;][]{ingels2011high}, a nationally representative dataset that follows the 2009 cohort of ninth graders in the United States, to estimate the effect of taking advanced mathematics coursework (AP/IB) on four-year college enrollment. The analytical sample contains $9,167$ students. The treatment is a binary indicator of whether a student has passed an AP/IB math exam at least once, and the outcome is whether a student has been enrolled in a four-year college. To reduce the risk of violating causal identification assumptions, we condition on $17$ covariates at both the student and school levels, including socioeconomic status, prior math achievement, school climate, and the proportion of twelfth graders at the school who have taken AP/IB coursework\footnote{There are 21 covariates in total after converting the categorical variables into dummy variables.}. The estimated propensity scores overlap well across the treated and control groups, which supports the positivity assumption (See Appendix~\ref{app:identification}). Because the individual effects are never observed in the empirical data, we assess the mentee model's faithfulness (the fine-tuned Gemma~E2B) against its mentor (the X-learner), and evaluate the external credibility based on agreement with the established literature.

\subsection{Results}

\begin{figure}[htbp]
    \centering
    \includegraphics[width=1.0\textwidth]{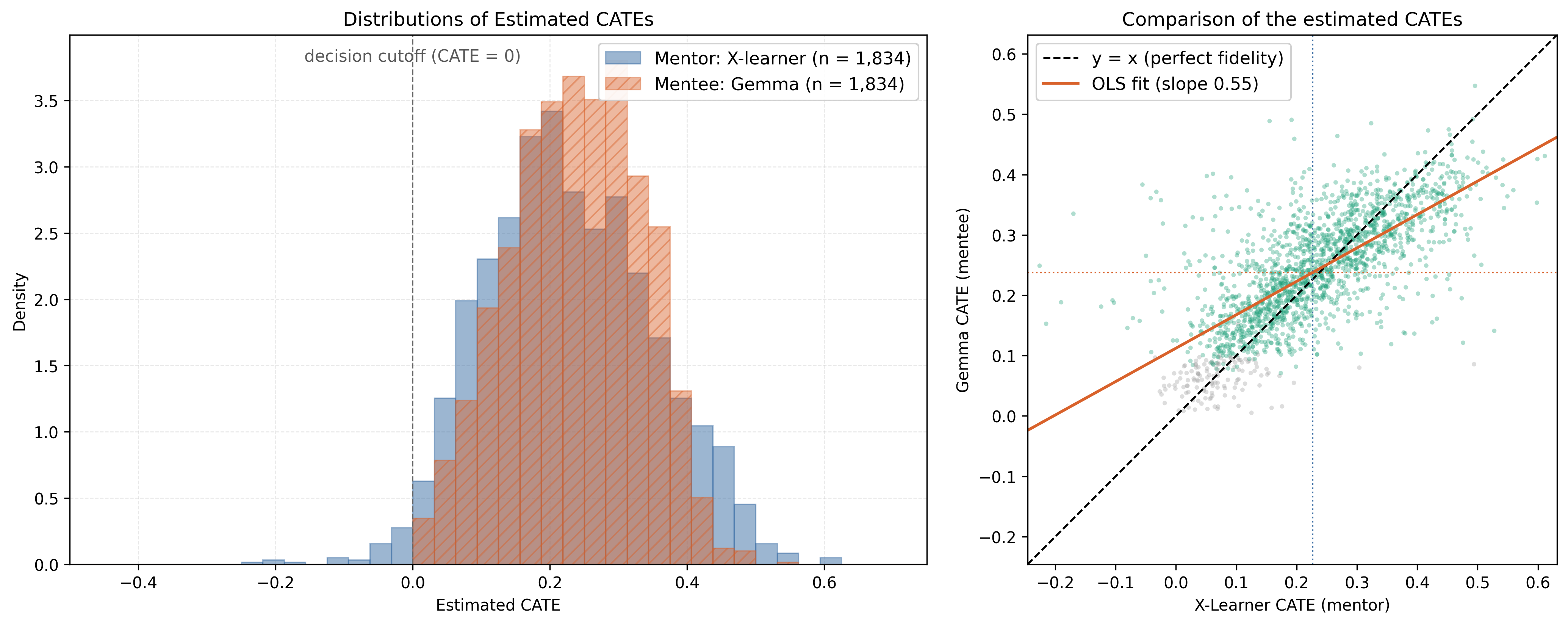}
    \caption{Distributions of the estimated CATEs and mentee fidelity on the held-out test set ($n = 1{,}834$). Left: densities of the CATEs estimated by the X-learner mentor (blue) and the fine-tuned Gemma mentee (orange, hatched), with the dashed line marking the decision cutoff at CATE $= 0$. Right: mentee-estimated CATEs plotted against mentor-estimated CATEs; the dashed line $y = x$ represents perfect fidelity, and the orange line is the ordinary least squares fit (slope $= 0.55$; Pearson $r = 0.71$).}
    \label{fig:emp_dist}
\end{figure}

Table~\ref{tab:emp_summary} summarizes all the key results. The X-learner estimates an overwhelmingly beneficial effect of advanced mathematics coursework, with an ATE of $0.230$ on the probability scale and a positive conditional effect for $98.3\%$ of students. The left panel in Figure~\ref{fig:emp_dist} shows the distribution to be unimodal and roughly symmetric about that average. The leading moderators are prior mathematics achievement and socioeconomic status, with fANOVA variance shares of $17\%$ and $13\%$, followed at a distance by educational expectations and school climate at $5\%$ and $4\%$. The estimated effect declines monotonically across quartiles of both leading moderators (Table~\ref{tab:emp_subgroup}), so the advanced mathematics benefit most the students who are less likely to enroll in college. This moderator structure and its direction are consistent with substantive conclusions of \textcite{byun2015advanced}, who studied the same question with propensity score matching on an earlier nationally representative cohort, the Education Longitudinal Study of 2002 \parencite[ELS:2002;][]{ingels2004education}, supporting the external credibility of the surface the mentee is asked to learn.

\begin{table}[!h]
\caption{Empirical results: mentor-level quantities (X-learner) and mentee fidelity (fine-tuned Gemma~E2B) on the test split. Decisions follow the sign rule to treat if $\hat\tau>0$. CATE = conditional average treatment effect (a probability risk difference).}
\label{tab:emp_summary}
\centering
\footnotesize
\begin{tabular}{l l}
\hline
\multicolumn{2}{l}{\emph{Mentor (X-learner)}}\\
Average treatment effect (risk difference) & $0.230$, 95\% CI $[0.208,\,0.258]$ \\
Conditional effects positive & $98.3\%$ \\
CATE distribution (mean / SD / median) & $0.230$ / $0.122$ / $0.227$ \\
Leading moderators (fANOVA variance share) & prior math $0.17$, SES $0.13$, expectations $0.05$, climate $0.04$ \\
First-order (additive) closure & $R^2=0.68$ \\
\quad $+$ detected interaction (math $\times$ climate) & $R^2=0.74$ \\
Decisions (test split) & \textsf{treat} $1{,}800$ $(98.1\%)$ / \textsf{do-not-treat} $34$ $(1.9\%)$ \\
\hline
\multicolumn{2}{l}{\emph{Mentee vs.\ mentor (test $n{=}1{,}834$)}}\\
Pearson $r$ / slope & $0.71$ / $0.55$ \\
Estimate range (mentee / mentor) & $[+0.005,\,+0.547]$ / $[-0.226,\,+0.612]$ \\
Decisions (test split) & \textsf{treat} $1{,}834$ $(100\%)$ / \textsf{do-not-treat} $0$ \\
Decision agreement & $0.981$ \\
\quad always-treat baseline & $0.981$ \\
\quad mentor \textsf{do-not-treat} recovered & $0$ of $34$ \\
Top-1 citation agreement / Intersection over union & $0.81$ / $0.77$ \\
Interaction-citation rate (mentee / mentor) & $0.16$ / $0.28$ \\
Narration grounding audit (full pass) & $0.988$ \\
Decomposition self-closure / uniqueness & $0.997$ / $0.999$ \\
parse rate & $1.00$ \\
\hline
\end{tabular}
\end{table}

A purely additive reconstruction from the first-order ALE terms closes $0.68$ of the variance in the mentor's estimates, and the data-driven screen selects prior mathematics achievement $\times$ school climate as the single retained interaction, which further lifts the decomposition closure to $0.74$. The second strongest pair adds only $2.3\%$. Therefore, this CATE surface from the X-learner is additive-dominant, which is consistent with \textcite{byun2015advanced}'s report of only minor interactions. Given that the X-learner flattens an effect surface toward apparent additivity as discussed in Section~\ref{sec:sim_results}, part of the remaining residual variance is likely noise rather than undetected structure. We read the empirical surface as additive-dominant with one modest but real interaction, and turn to what the mentee makes of it.

\begin{table}[!h]
\caption{Estimated effect by quartile of the two leading moderators (mentor X-learner; Q1 = lowest, Q4 = highest). The benefit is concentrated among the lowest-status students.}
\label{tab:emp_subgroup}
\centering
\footnotesize
\begin{tabular}{l c c c c c}
\hline
Quartile (low $\rightarrow$ high) & Q1 & Q2 & Q3 & Q4 & Q1$-$Q4 gap \\
\hline
Socioeconomic status & $+0.291$ & $+0.274$ & $+0.218$ & $+0.135$ & $+0.156$ \\
Prior mathematics achievement & $+0.310$ & $+0.242$ & $+0.198$ & $+0.169$ & $+0.141$ \\
\hline
\end{tabular}
\end{table}

As shown in the right panel in Figure~\ref{fig:emp_dist}, on the held-out test split of $1{,}834$ students, the mentee's estimates correlate with the mentor's at $r=0.71$, with a regression slope of $0.55$, which reflects the similar compression towards the mean from the simulation study. The mentor's estimates run from $-0.226$ to $+0.612$, whereas the mentee's run from $+0.005$ to $+0.547$, so that the entire negative tail of the mentor's surface is truncated in distillation. This is expected, since no more than 2\% of students have estimated negative effects from the X-learner. The mentee model does not have enough information to learn about the negative cases and therefore recommends all students to take the AP/IB courses, while the mentor recommends treating $1{,}800$ out of $1{,}834$ students. The lesson is that distillation may worsen the decisions in a severely imbalanced case, despite its nearly perfect accuracy. 

Attribution transmits far better than decisions. The mentee's top-cited covariates match the mentor's for $80.6\%$ of students, and about 77\% of the terms cited by either model are cited by both (a ratio of intersection over union at 0.77), so the explanation generally names the same reasons even when it misstates the size of the effect. It cites the mathematics-by-climate interaction for $16\%$ of students against the mentor's $28\%$, which mirrors the simulation's interaction-recovery gap under a realistic mentor. In addition, we evaluate the mentee narration's faithfulness by using the faithfulness audit of Section~\ref{sec: audit_gate} with no repair step. On the test set, $98.8\%$ of narrations pass in full, the stated decomposition sums to the stated effect for $99.7\%$ and $99.9\%$ of narrations are unique. The residual $1.2\%$ consists of minor slips, such as a direction word inconsistent with the sign of a small cited contribution, rather than hallucinated quantities. Narration quality survives distillation nearly intact.

The deployed mentee is the fine-tuned Gemma E2B carrying the LoRA adapter, which adds only 0.11 GB to the base weights. We built two interfaces over it, a dashboard in which a user sets a student's characteristics directly and a conversational assistant that accepts a plain-language description, and confirmed that both run end-to-end on an Apple M-series MacBook Pro. At 16-bit precision, the model occupies 10.3 GB on disk and inference peaks near 10.7 GB of memory, so a single estimate and its explanation are produced in about ten seconds with no network access of any kind, which means the student record never leaves the machine. That configuration needs 32 GB to sit alongside ordinary office software. For a 16 GB machine, we recommend the 4-bit variant, which occupies 3.6 GB and peaks near 4.1 GB at a negligible cost in fidelity. Appendix~\ref{app:size} provides the comparisons on the disk and memory usage among Gemma 4's variants. All source code and both deployment demonstrations are available at \url{https://github.com/cgpan/distillLA}.

\section{Discussion}

Across every comparison, the performance degradation is primarily caused by the mentor construction stage and is faithfully transmitted to the mentee. When fed with an oracle signal, the distillation step is nearly lossless for point estimation, distribution recovery, and key variable detection. The decomposition is candid since the first-order attributions explain nearly half the variance and recover most of the remainder after adding the interaction term. Whatever the retained terms cannot attribute, the mentee absorbs into its residual without fabricating a substitute from the noise variables. After switching from the oracle to an X-learner, performance degrades in all three as expected, since the X-learner shrinks the estimate toward the mean and flattens the surface toward additivity due to its finite-sample property. The mentee inherits both properties and generates the decomposition with a markedly higher first-order variance share than the oracle ($R^2$ of 0.73 against 0.50). The ALE-fANOVA averages estimation error across the covariate windows, so the narrated surface is closer to the truth than the X-learner's estimate is, and the mentee inherits that improvement as well. The practical implication is that the upstream model deserves careful scrutiny since it mostly influences what the mentee learns. 

One critical lesson is that fluency and faithfulness are distinct properties. In every condition we examine, the fine-tuned mentee models produce narrations that read well, cite covariates with correct calculation, and end with a decision stated in confident and actionable language. However, in the severely imbalanced simulation scenario, these seemingly reasonable narratives recommend actually harmful treatment to almost all students whom the treatment actually harms under the X-learner, although this issue is attenuated in the moderately imbalanced case. This danger can only be detected in a simulated scenario, where the ground truth is available through artificial design, and it remains a stubborn issue in any decision-making system. We believe this is worth particular attention in learning analytics, where generative technologies are arriving quickly, and the parents, students, counselors, or other stakeholders might be the least equipped to notice that a well-written explanation is wrong.

The pipeline has favorable properties for data privacy and offers several entry points for fairness constraints. Student records never leave the machine and are never transmitted to a third-party model provider. While the mentor construction stage requires a more capable LLM to generate the narration for the mentee to learn, this process can also run locally on an open-weight model like Gemma 4 32B or Alibaba's Qwen 3.8 35B on a MacBook Pro with 128 GB memory. To facilitate algorithmic fairness \parencite[]{mitchell2021algorithmic} in a prediction task, one may drop the sensitive variables (e.g, Race/ethinicty, gender, religion, etc.) from the pipeline entirely and keep the mentee blind to them \parencite[fairness through unawareness;][]{kusner2017counterfactual}, though whether omission actually produces fairer outcomes is contested and needs evaluation in context \parencite[]{dwork2012fairness, pan2024examining}. For a causal task, the tension is sharper since sensitive covariates are often needed to make the no-unmeasured-confounder assumption plausible (see Appendix~\ref{app:identification}) and cannot simply be dropped. One option is to keep them in estimation and remove them at the decision stage, such as regressing the estimated CATE onto the covariate subset that excludes sensitive variables. Another is to impose fairness constraints during estimation so that the resulting effects behave comparably across sensitive groups \parencite[]{suk2026fair, kim2023fair}. 

Deployment is where the pipeline earns its complexity. A stakeholder interacts with the deployed model in natural language and needs no facility with the X-learner, with ALE-fANOVA, or with the data format either of them expects. In contrast, conventional model deployment may require a fitted estimator, an interpretation tool, the input data in a rigid format, and typically a hosted LLM (e.g., ChatGPT, Claude) to take all the results and a pre-specified prompt to return the interpretation, which is both technically demanding and at risk of privacy disclosure to any third party. The proposed pipeline handles all these challenges securely and efficiently without compromising the student's privacy. The same design extends naturally to distill multiple upstream models into one mentee model. For example, in addition to the current capability for math course recommendation and interpretation, we can also distill a model for other STEM coursework recommendations, or ML models for dropout prediction, which makes the deployed artifact a general analysis assistant rather than a single-question tool. We leave a full evaluation of that extension to future work.

Several limitations are worth discussion. The simulation rests on five replications, and the empirical study uses one dataset, one treatment, and one outcome. Future work may generalize the conclusions by reproducing this pipeline with additional replications and across various ML or causal ML settings or datasets. The fine-tuned mentee compresses the magnitude in every condition, so individual point estimates should be read with caution even where correlation is high. LLMs are known to handle qualitative and linguistic material more reliably than precise decimal arithmetic \parencite[]{zhang2025comprehension, yang2025number}. One remedy to magnitude compression is to categorize the effect into several qualitative labels, such as largely beneficial, modestly beneficial, uncertain, moderately harmful, and largely harmful. We focus on the numerical effect to honestly expose the issue rather than conceal it behind a coarser scale. We also demonstrate that varying the mentee model size (from Gemma E2B to Gemma 26B) and precision (from INT4 to BF16) does not remove this issue. See Appendix~\ref{app:size} for comparisons among various Gemma variants. We also note that the decomposition describes statistical association within the fitted surface and does not represent a causal-path reading, so a narrated interaction should not be understood as a mechanism by which one covariate operates through another. Additionally, the mentee is an approximator of the mentor's effect surface rather than a causal estimator in our design since it never observes treatment or outcome and performs no causal identification.

\section{Conclusions}

We have presented a two-stage pipeline that distills an ML estimator, along with the fANOVA-based interpretation, into a small open-weight language model that predicts the effect and explains it in natural language, and that runs on commodity hardware without network access. The simulation design demonstrates that the mentee reproduces its mentor almost exactly at the oracle ceiling and inherits its mentor's flaws under a realistic estimator rather than hiding them or fabricating information. In the application on a nationally representative dataset, the pipeline recovers a substantively credible finding that advanced mathematics coursework benefits most the students least likely to enroll in college, and delivers it as an individualized decomposition that names the dominant moderators and the retained interaction without fabricating quantities. 

Two caveats are worth attention. First, fluency is cheap with fine-tuned LLMs since a distilled model writes equally well whether its mentor signal is exact or noisy, so that narration quality cannot be used as evidence that a narration is correct. Second, decisions are fragile in a severely imbalanced setting, where one recommendation is overwhelmingly common. Both the estimator and the mentee reproduce a rule that recommends it for nearly everyone. Reporting decision accuracy against the majority baseline or relying on the confidence interval of estimates for decision-making are the minimum safeguards we would attach to any deployment of this design.

\printbibliography

\appendix
\renewcommand{\thesection}{Appendix \Alph{section}}

\setcounter{secnumdepth}{1}
\setcounter{equation}{0}
\setcounter{figure}{0}
\setcounter{table}{0}
\setcounter{section}{0}


\newpage
\section{Causal Assumptions and Identification} \label{app:identification}

In the causal ML case study discussed in the main text, we define the target parameter, CATE, as $\tau(x) = \mathbb{E}[Y^{*}(1) - Y^{*}(0) \mid X = x]$. Because each student is observed under only one treatment condition, the individual-level potential outcome difference is never available \parencite[]{holland1986statistics}, and identification of $\tau(x)$ from the observed data $(X, A, Y)$ requires three assumptions:
 
\begin{itemize}
\item[(A1)] \textit{Consistency}: $Y_i = Y_i^{*}(A_i)$.
\item[(A2)] \textit{Conditional ignorability}: $\{Y^{*}(0), Y^{*}(1)\} \perp A \mid X$.
\item[(A3)] \textit{Positivity}: $0 < \Pr(A = 1 \mid X = x) < 1$ for all $x \in \mathcal{X}$.
\end{itemize}
 
Assumption (A1) states that the observed outcome is the potential outcome under the treatment actually received, which presumes no interference between students and a single well-specified version of the treatment \parencite[]{rubin1986comment}. For example, a student's college enrollment depends only on whether that student took advanced mathematics, not on how many classmates did, and an AP and an IB mathematics course count as the same intervention rather than two different ones. Assumption (A2) states that, conditional on the observed covariates, treatment assignment is independent of the potential outcomes, so that $X$ contains every factor jointly influencing treatment and the outcome \parencite[]{rosenbaum1983central}, which is also referred to as no unmeasured confounding or unconfoundedness. Assumption (A3) requires both treatment conditions to have positive probability at every covariate profile \parencite[]{rubin1974estimating}.
 
Under (A1) through (A3), $\tau(x)$ can be written in terms of observed data as
\begin{align}
\tau(x) &= \mathbb{E}[Y^{*}(1) \mid X = x] - \mathbb{E}[Y^{*}(0) \mid X = x] \notag \\
        &= \mathbb{E}[Y^{*}(1) \mid A = 1, X = x] - \mathbb{E}[Y^{*}(0) \mid A = 0, X = x]
           && \text{(conditional ignorability)} \notag \\
        &= \mathbb{E}[Y \mid A = 1, X = x] - \mathbb{E}[Y \mid A = 0, X = x]
           && \text{(consistency),}
\label{eq:identification}
\end{align}
with positivity ensuring that both conditioning events occur with positive probability. The right-hand side relies only on observable quantities and is what the estimators discussed in Section~2.3 approximate. By averaging over the covariate distribution, we obtain the ATE,
\begin{equation}
\tau_{\mathrm{ATE}} = \mathbb{E}\!\left[Y^{*}(1) - Y^{*}(0)\right]
 = \mathbb{E}_{X}\!\left[\tau(X)\right]
 = \int_{\mathcal{X}} \tau(x)\, dF_{X}(x),
\label{eq:ate}
\end{equation}
which is estimated by the sample analog $n^{-1}\sum_{i=1}^{n} \hat{\tau}(X_i)$. This is the quantity each narration reports as the average effect, since the grand mean of the fANOVA decomposition of $\hat{\tau}(x)$ coincides with it.

\subsection{Evaluate the causal assumptions in empirical analysis }

\begin{figure}[htbp]
    \centering
    \includegraphics[width=0.9\textwidth]{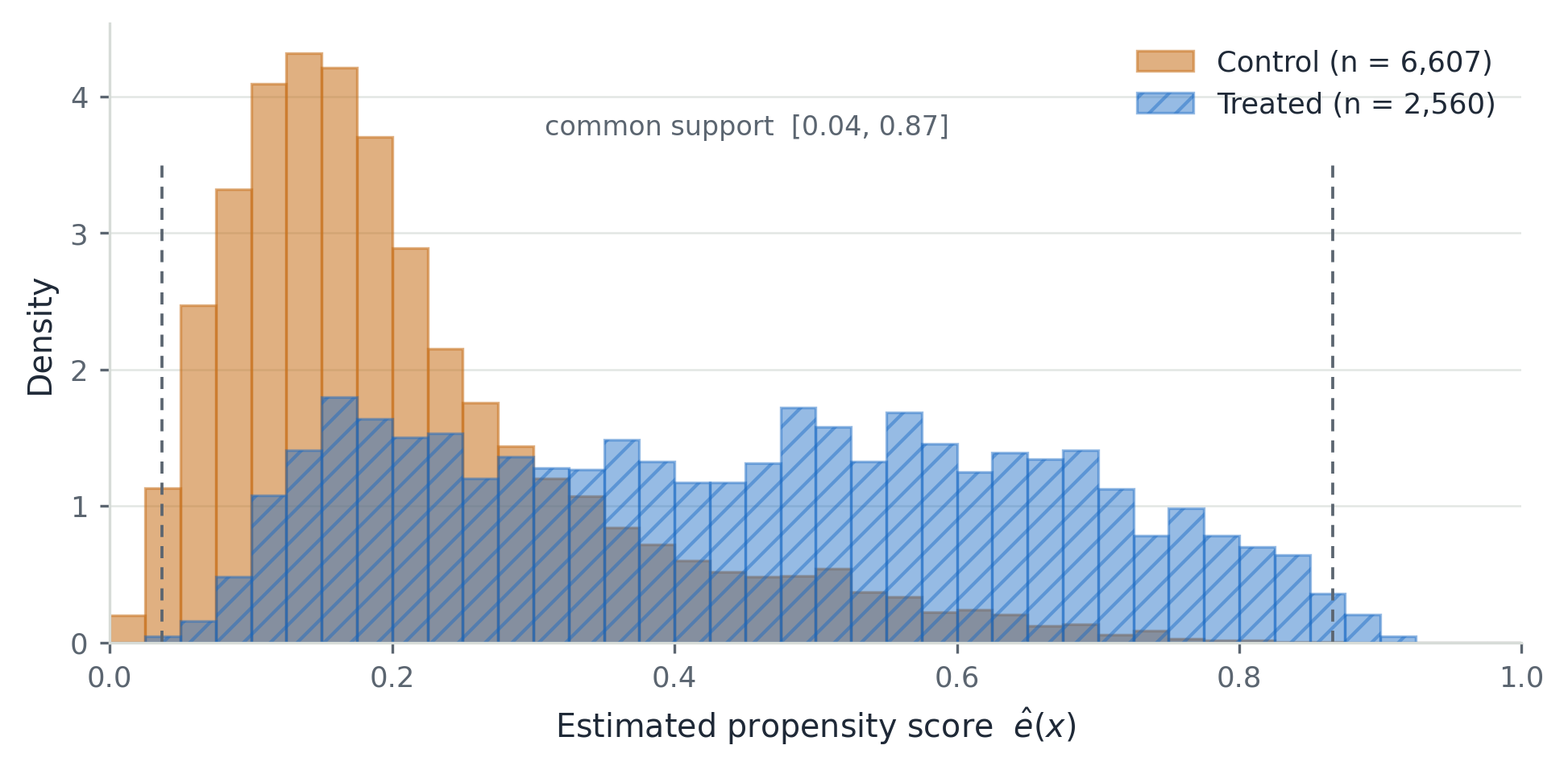}
    \caption{Distribution of cross-fitted propensity scores by treatment group for the HSLS analytical sample. Dashed lines mark the bounds of common support.}
    \label{fig:pscores}
\end{figure}

As a common challenge for causal inference in educational empirical data, students from the same classroom or school may share common features, and their choice of treatment might affect each other's. The treatment variable considers AP and IB courses as a single unified one, so (A1) holds only approximately. Assumption (A2) is untestable. We include 17 covariates from both student- and school-level to lower the risk of violation. A sensitivity analysis reporting how strong an unmeasured confounder would need to be in order to overturn the estimates is a natural extension of this work. We do not include this analysis here since we use the causal ML as a case study to demonstrate the proposed pipeline. Assumption (A3) is the only one that can be tested using empirical data. It is well supported for the great majority of the analytical sample. As shown in Figure~\ref{fig:pscores}, the treated and control distributions share common support over $[0.04,,0.87]$, and only $1.2\%$ of the sample ($107$ of $9{,}167$) falls outside that region.

These assumptions attach to the upstream estimator (i.e., X-learner) rather than to the mentee model. The mentee observes neither $A$ nor $Y$ and performs no identification of its own, so its output carries a causal reading only when three causal assumptions hold and the mentor both estimates $\tau(x)$ well and is faithfully approximated. In addition, the fANOVA decomposition describes how the fitted surface varies with the covariates, so a narrated first-order or second-order term is a feature of that surface rather than a causal pathway.


\newpage
\section{Worked Example: One Student Through Both Stages} \label{app:traces}

To better clarify how the proposed pipeline works, we provide an example of a single student to show how the personal data flows and what the LLM's traces and narration look like. All text is reproduced from the empirical run. The student is drawn from the validation split, so the mentor authored a trace for this student, but the student contributed no gradient step to LoRA training, and the mentee output shown below is generated rather than memorized. For demonstration purpose, we keep sensitive variables (e.g., the gender, race/ethnicity) in the pipeline, while practitioners might want to remove them in the real school deployment.
 
The example also documents the training-example format referenced in the Methods section. Listing~\ref{lst:sys-mentee} is the fixed system prompt, Listing~\ref{lst:covariates} is the user prompt, and Listing~\ref{lst:mentor-trace} is the assistant target over which the token-level cross-entropy is computed.

\begin{table}[htbp]
\centering
\caption{Mentor and Mentee decompositions for the same student.}
\label{tab:worked-compare}
\footnotesize
\begin{tabular}{@{} l r r @{}}
\toprule
Term & Mentor & Mentee \\
\midrule
Baseline (grand mean)                    & $+0.2289$ & $+0.2289$ \\
School control                           & $-0.0463$ & $-0.0463$ \\
Socioeconomic status                     & $+0.0382$ & $+0.0383$ \\
Student educational expectation          & $+0.0306$ & $+0.0306$ \\
Mathematics achievement                  & $-0.0409$ & not cited \\
Student sex                              & not cited & $+0.0147$ \\
Mathematics achievement $\times$ school climate & $-0.0156$ & $+0.0731$ \\
Residual                                 & $+0.0201$ & $-0.0101$ \\
\midrule
Estimated effect                         & $+0.2150$ & $+0.3292$ \\
Decision                                 & treat     & treat \\
\bottomrule
\end{tabular}
 
\vspace{0.5em}
\begin{minipage}{0.92\textwidth}
\raggedright
\footnotesize
\textit{Note.} The outcome is the probability scale of four-year college enrollment. Both decompositions close arithmetically. The interaction term carries the opposite sign, and the mentee's narration describes it as amplifying the benefit.
\par
\end{minipage}
\end{table}
 
\subsection*{Stage 1: Mentor Construction}
 
The X-learner estimate for this student is $\hat{\tau} = +0.2150$ on the probability scale of four-year college enrollment. The fANOVA decomposition cites four first-order terms and one second-order term. The full ledger appears in the left column of Table~\ref{tab:worked-compare}.
 
The covariate block in Listing~\ref{lst:covariates} is the only student-specific text that ever reaches the mentee. It carries no estimate, no decomposition, and no decision.
 
\begin{lstlisting}[caption={Student covariate block (user prompt at both stages)}, label={lst:covariates}]
X1 Student's sex: Male
X1 Student's race/ethnicity-composite: Hispanic, race specified
X1 Student dual-first language indicator: First language is a non-English language only
X1 Mathematics standardized theta score: +70.410
X1 Socio-economic status composite: -0.373
X1 Parents'/guardians' highest level of education: High school diploma or GED
X1 How far in school 9th grader thinks he/she will get: Complete a Bachelor's degree
X1 How far in school parent thinks 9th grader will go: Complete a Master's degree
X1 Mathematics assessment accommodations: No
X1 Math teacher's race/ethnicity-composite: Asian, non-Hispanic
X1 School control: Catholic or other private
X1 School locale (urbanicity): City
X1 School geographic region: Midwest
X1 Scale of administrator's assessment of school climate: -0.850
X1 School staff expectations (counselor-reported average): +0.620
X1 School offers AP mathematics: 1
C2 Percent of 12th graders who have taken AP course(s): +50.000
\end{lstlisting}
 
Listing~\ref{lst:evidence} shows the prompt for the decomposition passed to the trace-authoring LLM as structured evidence. Every number the narration is permitted to use appears here, and the \texttt{context} clauses are the phrase pool entries described in the Methods section (See Table~1). Note that the two categorical drivers carry no \texttt{context} clause, since a monotone tendency is not defined for them.
 
\begin{lstlisting}[caption={Trace-authoring user prompt (the mentor's decomposition evidence)}, label={lst:evidence}]
------ STUDENT DECOMPOSITION ------
predicted_effect = +0.2150
average effect across all students (ATE) = +0.2289
residual = +0.0201 ; confidence = high ; decision = treat
cited drivers:
  - X1 School control: observed value 'Catholic or other private';
    attributed contribution -0.0463 -> LOWERS
  - mathematics standardized theta score: observed value 70.41 (above average);
    attributed contribution -0.0409 -> LOWERS; context: in line with the model's
    general pattern that lower mathematics standardized theta score is associated
    with larger benefits
  - socio-economic status composite: observed value -0.373 (below average);
    attributed contribution +0.0382 -> RAISES; context: in line with the model's
    general pattern that lower socio-economic status composite is associated
    with larger benefits
  - X1 How far in school 9th grader thinks he/she will get: observed value
    'Complete a Bachelor's degree'; attributed contribution +0.0306 -> RAISES
  - INTERACTION mathematics standardized theta score x school climate scale
    (administrator-assessed): observed values 70.41 (above average) and -0.85
    (below average); attributed contribution -0.0156 -> DAMPENS; context: in line
    with the model's finding of a joint effect between mathematics standardized
    theta score and school climate scale beyond their separate contributions
Write the explanation and emit the JSON.
\end{lstlisting}
 
The accepted trace in Listing~\ref{lst:mentor-trace} is the LoRA training target. It opens with a scratchpad that states the arithmetic explicitly, followed by the structured output. The \texttt{rationale} field is reproduced separately below for readability.
 
\begin{lstlisting}[caption={Accepted mentor trace (assistant target for LoRA training)}, label={lst:mentor-trace}]
<think>
baseline +0.2289 + X1CONTROL -0.0463 + X1TXMTSCOR -0.0409 + X1SES +0.0382
+ X1STUEDEXPCT +0.0306 + X1TXMTSCORxX1SCHOOLCLI -0.0156 + residual +0.0201
= predicted effect +0.2150
sign +; confidence high -> decision treat
</think>
{"predicted_cate":0.215,"baseline":0.2289,
 "cited_terms":[
   {"order":1,"features":["X1CONTROL"],"values":["Catholic or other private"],
    "standings":[null],"contribution":-0.0463},
   {"order":1,"features":["X1TXMTSCOR"],"values":[70.41],
    "standings":["above average"],"contribution":-0.0409},
   {"order":1,"features":["X1SES"],"values":[-0.373],
    "standings":["below average"],"contribution":0.0382},
   {"order":1,"features":["X1STUEDEXPCT"],"values":["Complete a Bachelor's degree"],
    "standings":[null],"contribution":0.0306},
   {"order":2,"features":["X1TXMTSCOR","X1SCHOOLCLI"],"values":[70.41,-0.85],
    "standings":["above average","below average"],"contribution":-0.0156,
    "direction":"dampens"}],
 "residual":0.0201,"confidence":"high","decision":"treat",
 "rationale":"<reproduced below>"}
\end{lstlisting}
 
\noindent\textbf{Mentor narration.}
\begin{quote}
\itshape\small
This student's predicted effect of $+0.2150$ is slightly below the average effect of $+0.2289$. The strongest downward driver is school control, observed as ``Catholic or other private,'' which lowers the benefit by $-0.0463$; the mathematics standardized theta score, observed at $70.41$ (above average), further lowers the benefit by $-0.0409$, in line with the model's general pattern that lower mathematics standardized theta score is associated with larger benefits. Conversely, the socio-economic status composite, observed at $-0.373$ (below average), raises the benefit by $+0.0382$, in line with the model's general pattern that lower socio-economic status composite is associated with larger benefits, and the student's expectation ``Complete a Bachelor's degree'' adds $+0.0306$. The interaction between mathematics standardized theta score ($70.41$, above average) and school climate scale ($-0.85$, below average) dampens the effect by $-0.0156$, in line with the model's finding of a joint effect beyond their separate contributions. Because the effect is clearly distinguishable from zero, the recommendation is to enroll the student in AP/IB math coursework.
\par
\end{quote}
 
\subsection*{Stage 2: Mentee Deployment}
 
At deployment the mentee receives the fixed system prompt in Listing~\ref{lst:sys-mentee} and, as the user turn, exactly the covariate block of Listing~\ref{lst:covariates}. It has no access to the estimator, or the decomposition. Everything it emits is produced in a single forward pass from the covariates alone.
 
\begin{lstlisting}[caption={Mentee system prompt (fixed across all students)}, label={lst:sys-mentee}]
You are a careful causal-inference analyst. Given ONE student's covariates,
estimate the treatment effect of taking AP/IB math coursework on 4-year college
enrollment (a probability risk difference) and explain it in faithful,
individualized prose that cites each key driver's observed value and its
contribution, compares to the average effect, notes any interaction, and ends
with the confidence and decision. Emit one JSON object.
\end{lstlisting}
 
\begin{lstlisting}[caption={Mentee output, generated in one pass (verbatim, scratchpad only)}, label={lst:mentee-trace}]
<think>
baseline +0.2289 + X1TXMTSCORxX1SCHOOLCLI +0.0731 + X1CONTROL -0.0463
+ X1SES +0.0383 + X1STUEDEXPCT +0.0306 + X1SEX +0.0147 + residual -0.0101
= predicted effect +0.3292
sign +; confidence high -> decision treat
</think>
\end{lstlisting}
 
\noindent\textbf{Mentee narration.}
\begin{quote}
\itshape\small
This student's predicted effect of $+0.3292$ is notably higher than the average effect of $+0.2289$ across all students. The strongest driver is the interaction between mathematics standardized theta score (observed at $70.41$, above average) and school climate scale (observed at $-0.85$, below average), which amplifies the benefit by $+0.0731$, in line with the model's finding of a joint effect beyond their separate contributions; however, attending a ``Catholic or other private'' school lowers the benefit by $-0.0463$, while a socio-economic status composite of $-0.373$ (below average) raises it by $+0.0383$, consistent with the model's pattern that lower socio-economic status is associated with larger benefits. Additionally, the student's expectation to ``Complete a Bachelor's degree'' raises the effect by $+0.0306$, and being ``Male'' adds $+0.0147$. Because the effect is clearly distinguishable from zero, the recommendation is to enroll the student in AP/IB math coursework.
\par
\end{quote}
 
\subsection*{What Transfers and What Does Not}
 
Table~\ref{tab:worked-compare} places the two ledgers side by side. The comparison is instructive precisely because it is mixed. The output format is exact, including the scratchpad, the JSON schema, and the closing decision sentence. The arithmetic closes in both traces without a repair step. Three of the mentor's five cited terms are reproduced to within $0.0001$. The mentee identifies the same interacting pair the data-driven screen selected upstream, which it could only have learned from the traces, since nothing in the covariate block marks that pair as special.
 
The point estimate is $+0.1142$ above the mentor's, which places this student in the upper tail of the mentor--mentee deviation distribution rather than at its center. The interaction is reported with the opposite sign and roughly $4.7$ times the magnitude, so a term the estimator found to dampen this student's benefit is narrated as the single largest source of amplification. And the citation set has two disagreements. The mentee drops mathematics achievement, which is the mentor's second-strongest driver, and substitutes student sex, a covariate that passes the global eligibility filter but that the mentor did not cite for this student.

\newpage

\section{Does the mentee's size or precision remove magnitude compression?} \label{app:size}

The realistic-mentor X-learner compresses large treatment effects toward the average, as evidenced by a calibration slope near $0.50$ (Table~\ref{tab:sim_richer}). We therefore conduct another experiment to compare the Gemma 4 variants' performance in estimate compression, disk usage, peak memory, inference (token-generation) speed across different model sizes from E2B to 26B, and precisions of 4-bit and BF16. The experiment uses the same DGP from the simulation study.

\begin{table}[!h]
\caption{Mentee size and precision sweep under the X-learner and the oracle mentor. Disk and peak inference memory in GB. Throughput in tokens per second. Here, E denotes Effective/Edge architectures optimized for low-memory deployment, where the total parameter count exceeds the active parameter count. MoE indicates a sparse Mixture-of-Experts model.}
\label{app:tab:size}
\centering
\footnotesize
\begin{tabular}{l c c c c c c}
\hline
Mentee & Precision & Disk & Peak infer. & Tokens/s & Pearson $r$ & Slope\\
\hline
\multicolumn{7}{l}{\emph{X-learner mentor}}\\
Gemma E2B          & 4-bit & 3.6  & 4.1  & 1184 & 0.880 & 0.759\\
Gemma E2B          & BF16  & 10.3 & 10.7 & 861  & 0.887 & 0.765\\
Gemma E4B          & 4-bit & 5.3  & 6.8  & 686  & 0.887 & 0.757\\
Gemma E4B          & BF16  & 16.0 & 17.5 & 493  & 0.891 & 0.759\\
Gemma 26B (MoE)    & 4-bit & 15.6 & 30.6 & 316  & 0.895 & 0.763\\
\hline
\multicolumn{7}{l}{\emph{Oracle mentor, same mentees}}\\
Gemma E2B          & BF16  & 10.3 & 10.7 & 789  & 0.966 & 0.960\\
Gemma 26B (MoE)    & 4-bit & 15.6 & 30.6 & 305  & 0.996 & 0.987\\
\hline
\end{tabular}
\end{table}

Table~\ref{app:tab:size} shows that the magnitude compression is a common issue for all variants, and the difference is trivial under the realistic estimator X-learner. A bigger and higher-precision Gemma model cannot remedy it. By comparing the results between the oracle mentor and the X-learner mentor, we can conclude that the magnitude compression comes mostly from the X-learner, and the distillation process honestly inherits this issue from the realistic X-learner. We note that magnitude compression is a finite-sample property of the X-learner, which might disappear as sample size increases.

Since the Gemma variants have similar performance under the realistic learner, the cheapest E2B Gemma with 4-bit is an optimal choice, which
occupies $3.6$~GB on disk and peaks near $4.1$~GB of memory, comfortably inside a consumer laptop.   

\end{document}